\documentclass{article}
\usepackage{lipsum}
\usepackage{authblk}
\usepackage[english]{babel}
\usepackage{amsfonts}
\usepackage{amsmath}
\usepackage[top=2cm, bottom=2cm, left=2cm, right=2cm]{geometry}
\usepackage{fancyhdr}
\usepackage{multicol}
\usepackage{epsfig}
\usepackage{graphics}
\usepackage{epsfig}
\usepackage{lipsum}
\usepackage{fancyhdr}
\usepackage{cite}
\usepackage[table]{xcolor}
\usepackage{multirow}
\renewenvironment{abstract}{
	
	\hfill\begin{minipage}{0.95\textwidth}
		\rule{\textwidth}{1pt}}
		{\par\noindent\rule{\textwidth}{1pt}\end{minipage}
	}

\newcommand{\bra}[1]{\bigl\langle #1 \bigr|}
\newcommand{\ket}[1]{\bigl| #1 \bigr\rangle}

\makeatletter
\makeatletter
\newcommand{\row}[1]
{\mathord{\buildrel{\lower3pt
\hbox{$\scriptscriptstyle\rightarrow$}}\over #1}}

\newcommand{\dyadic}[1]{\mathord{\dyadic@rrow{#1}}}
\newcommand{\dyadic@rrow}[1]{
\begin{picture}(12,12)(-1,0)
\put(-3,12){\makebox(0,0)[t]{$\scriptscriptstyle\downarrow$}}
\put(-3,13){\makebox(0,0)[l]{$\scriptscriptstyle\longrightarrow$}}
\put(5,0){\makebox(0,0)[b]{$#1$}}
\end{picture}
}
\makeatother

\begin{document}
\title{\textbf{Memory effects on bidirectional Teleportation}}
\author[1,2]{ \textbf{C. Seida}}
\author[2]{ \textbf{S. Seddik}}
\author[1]{ \textbf{Y. Hassouni}}
\author[2]{ \textbf{A. El Allati}}
\affil[1]{\small ESMaR, Faculty of Sciences, Mohammed V University in Rabat, Morocco}
\affil[2]{\small Laboratory of R\&D in Engineering Sciences, Faculty of Sciences and Techniques Al-Hoceima, Abdelmalek Essaadi University, Tetouan,
	Morocco}

\maketitle

\begin{abstract}
In this contribution, we have investigated the bidirectional quantum teleportation $(BQT)$ of single qubit states using a Bell state influenced by decoherence channels with memory, dephasing and amplitude damping channels. The expressions of the negativity,  as a measure of the entanglement remaining in the $BQT$ quantum channel, the teleportation fidelities and the quantum Fisher information are also evaluated. We find that both these last quantities depend on the survival amount of entanglement in the $BQT$ quantum channel, on the decoherence factor and on the correlation degree of the decoherence channel. We show that in the Markovian regime, the Negativity, the teleportation average fidelities and the quantum Fisher information are slightly enhanced by considering the classical channel correlations. Besides, in the non-Markovian regime, these three quantities could be improved for a long period of time.
\end{abstract}
\section{Introduction}
Quantum teleportation $(QT)$ provides a complete transmission of an unknown quantum information from a sender to a distant receiver. In $1993$,  physicist $Bennet$ $et$ $al.$ \cite{1} proposed the concept of quantum teleportation. In this seminal work, $QT$  is a one-way transmission of an unknown quantum information. In other words, only Alice has the ability to transmit her unknown quantum state to Bob. Indeed, the key ingredient of $QT$ is sharing an entangled Bell state and classical communication with local operations $(LOCC)$ \cite{a}. The $QT$ can be used to link several quantum computers that form a quantum network \cite{2,allati1}, It is also a cornerstone for the development of quantum information technologies such as quantum repeaters \cite{3} and universal computing \cite{4}. Accordingly, $QT$ has become a research hot-spot and has been investigated theoretically and experimentally \cite{b,c,Lev0}.  In fact, the first proposal for bidirectional teleportation is given by $Vaidman$ \cite{Lev}. Later on, $Kiktenko$ $et$ $al.$  \cite{5} addressed the issue of the extension of the standard teleportation scheme suggested in Ref \cite{1} in order to allow two legitimate users to exchange their quantum states using the same quantum resource and two classical communication channels. Successfully, numerous modified $BQT$ schemes  have been designed by means of different quantum resources \cite{6,7}. The accuracy of these protocols is usually quantified with the teleportation fidelities \cite{7a} and the quantum Fisher information \cite{7b}. The teleportation fidelity quantifies the teleportation accuracy of the whole qubit state. However, the teleportation of information that is encoded in the weight parameter of the qubit instead of the whole state is quantified by means of the quantum Fisher information (QFI)\cite{d}. Beside quantum teleportation, $QFI$ is critical for parameter estimation, with a large QFI value corresponding to a high degree of parameter estimation \cite{9, 10}.
\\

As a matter of fact,  decoherence is a major obstacle to achieve quantum information processing $(QIP)$, such as quantum teleportation and quantum computing tasks \cite{1,111}. The undesirable contact of an open quantum system with its surrounding environment induces decoherence and causes the degradation of the teleportation fidelity \cite{0}, the entanglement \cite{00} and the quantum coherence \cite{000}. The investigation of this issue has been extended to the theory of open quantum systems \cite{12}, which has attracted the interest of researchers in recent years. Due to the importance of preserving quantum resources and overcoming decoherence to accomplish quantum information tasks with high performance, several methods have been proposed in the literature, such as the purification \cite{13}, the quantum zeno effect \cite{15} and the weak and reversal measurements \cite{16,17}. Apart from these methods, once an open quantum system interacts with its surrounding environment, it consistently loses information to the environment. In this case, decoherence induces the Markovian dynamics \cite{177,12}. However, in the non-Markovian dynamics, the open system both loses and retrieves information from the surrounding environment \cite{18}.\\ 

In fact, there are two types of memory effects that are related to the information flow back from the environment to the system \cite{19}. The first type is due to the classical correlations between the successive actions of the noisy environment on the quantum system \cite{20}, and the second type is  due to the temporal correlations arising during the system's evolution \cite{199}. The interactions between the system of interest and its environment can be described by quantum channels. Indeed, quantum channels, with or without memory,  indicate that their successive actions are respectively correlated to or independent from each other. The investigations of the classical and temporal memory effects are separately discussed. The exploitation of the classical correlations, between consecutive actions of the quantum channel, may increase the quantum coherence \cite{21}, protect the $QFI$ from depolarizing and phase-flip noise channels \cite{211} and enhance the $BQT$ of information \cite{PD0}. On the other hand, the non-Markovianity memory effect could efficiently suppress the decoherence effect \cite{212}. The interplay between classical correlated actions of quantum channels and non-Markovianity is discussed in \cite{24}. However, the combined classical and temporal correlation effect on quantum coherence is investigated in \cite{241} and on unidirectional teleportation  is discussed in \cite{22,23}.\\

In this paper, we evaluate  the effects of combined memory types on bidirectional quantum teleportation, where the quantum channel of the $BQT$ is distributed via typical correlated decoherence channels, such as a dephasing channel and an amplitude damping channel. We show that the entanglement of the quantum channel of the $BQT$, the average teleportation fidelities and the quantum Fisher information could be further enhanced for a large range of time by taking into account the combined effects of the classical and temporal correlations between the successive actions of the decoherence channels.\\

This paper is organized as follows: Sec. 2 discusses the $BQT$ procedure and presents entanglement and BQT teleportation quantifiers such as negativity $\mathcal{N}^{(PT)}$, teleportation average fidelities, and quantum Fisher information with respect to the weight parameter. In Sec. 3, we evaluate the analytical expressions of $\mathcal{N}^{(PT)}$, the teleportation fidelities, and the quantum Fisher information for the correlated dephasing and the amplitude damping channels. We summarize  the findings of this paper in Sec. 4.

\section{Bidirectional quantum teleportation}
In this section, we review the $BQT$ protocol presented in Ref \cite{5,d}. We also present the negativity of the partially transposed density operator, which is an entanglement measure. The average fidelity of teleportation as well as the teleported quantum Fisher information are given to estimate the behavior of the bidirectional quantum teleportation.
\subsection{Bidirectional teleportation protocol}
 The primary objective of this protocol is to bilaterally teleport the quantum states $\ket{\mathcal{S}_A}$ and $\ket{\mathcal{S}_B}$ between two users. We assume that Alice and Bob are handed two different states, $\ket{\mathcal{S}_A}$ and $\ket{\mathcal{S}_B}$ \cite{5,d}:
\begin{equation}\label{Qubits}
    \ket{\mathcal{S_A}}=\cos({\frac{\vartheta_A}{2}})\ket{0}_A+ e^{i\varphi_A}\sin({\frac{\vartheta_A}{2}})\ket{1}_A
    ,\quad
    \ket{\mathcal{S_B}}=\cos({\frac{\vartheta_B}{2}})\ket{0}_B + e^{i\varphi_B}\sin({\frac{\vartheta_B}{2}})\ket{1}_B.
\end{equation}
In addition, Alice and Bob are given two auxiliary qubits which play the role of Trigger qubits, given as :
\begin{equation}\label{triggers}
\ket{\mathcal{T_A}}=\cos\Bigr(\frac{\tilde\vartheta_A}{2}\Bigl)\ket{0}_A+\sin\Bigr(\frac{\tilde\vartheta_A}{2}\Bigl)\ket{1}_A, \quad
\ket{\mathcal{T_B}}=\cos\Bigr(\frac{\tilde\vartheta_B}{2}\Bigl)\ket{0}_B+\sin\Bigr(\frac{\tilde\vartheta_B}{2}\Bigl)\ket{1}_B,
\end{equation}
whereas: $\vartheta_{i} \in [0,\pi]$, $\varphi_{i} \in [0,2\pi]$ and $\tilde\vartheta_{i} \in [0,\pi]$, for $(i=A,B)$.
\\

Since we are interested in quantifying the accuracy of the $ BQT$  of the weight parameter by means of the quantum Fisher information,  we rewrite the states in Eqs.(\ref{Qubits}) and (\ref{triggers}) in the density operator representation. Hence, the states become:
\begin{eqnarray}{\label{qubits}}
\hat\varrho_{\mathcal{S}}^{(Alice)}=\frac{1}{2}\Bigr(\mathbf{I}+\sum_{{n=1,2,3}}\alpha_n\hat{\gamma}_n^{(A)}\Bigl), \quad \hat\varrho_{\mathcal{S}}^{(Bob)}=\frac{1}{2}\Bigr(\mathbf{I}+\sum_{n=1,2,3}\beta_n\hat{\delta}_n^{(B)}\Bigl),
\end{eqnarray}
where $\alpha_n$ and $\beta_n$, for $(n=1,2,3)$, are the Bloch vectors of Alice's and Bob's quantum states:  
\begin{eqnarray}
\alpha_n=\operatorname{Tr}\{\hat\varrho_{\mathcal{S}}^{(Alice)}\hat{\gamma}_n\} \quad\hbox{ and } \quad\quad  \beta_n=\operatorname{Tr}\{\hat\varrho_{\mathcal{S}}^{(Bob)}\hat{\delta}_n\} .
\end{eqnarray}
The operators  $\hat{\gamma_n}$ and $\hat{\delta_n}$ (for $n=1,2,3$) are the  Pauli matrices for Alice's and Bob's quantum states, respectively. 
Similarly, the trigger qubits could be written in the Bloch representation as:
\begin{equation}\label{trig}
\hat\varrho_{\mathcal{T}}^{(A)}=\frac{1}{2}\Bigr(\mathbf{I}+\sum_{n=1,2,3} \upsilon_n^{(A)}\hat{\Upsilon}_n^{(A)} \Bigl),\quad
\hat\varrho_{\mathcal{T}}^{(B)}=\frac{1}{2}\Bigr(\mathbf{I}+\sum_{n=1,2,3}\upsilon_n^{(B)}\hat{\Upsilon}_n^{(B)}\Bigl),
\end{equation}
where $\upsilon_n^{(A)}$, $\upsilon_n^{(B)}$ are the Bloch vectors and $\hat{\Upsilon}_n$ are the Pauli operators.
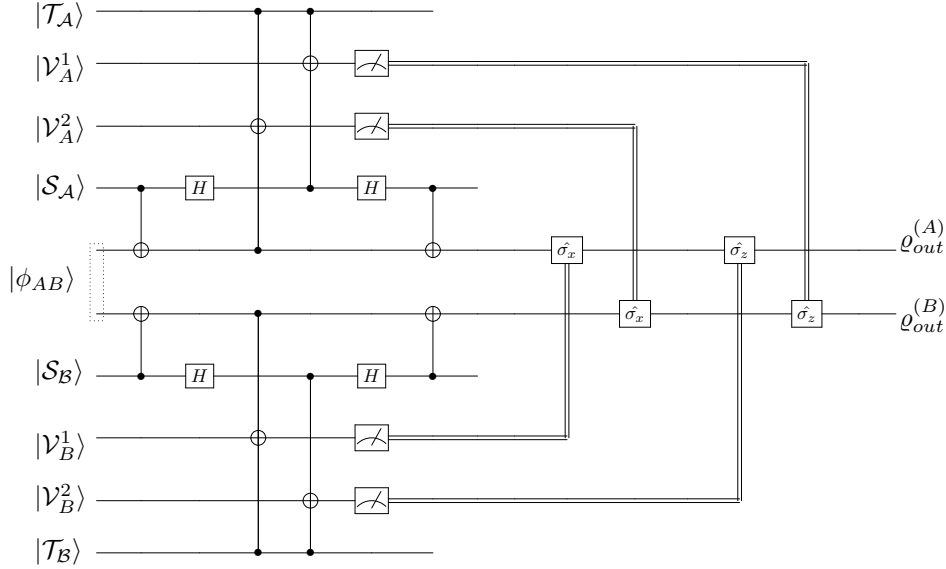
\begin{figure}[!htb]
\begin{center}
\includegraphics[scale=.7]{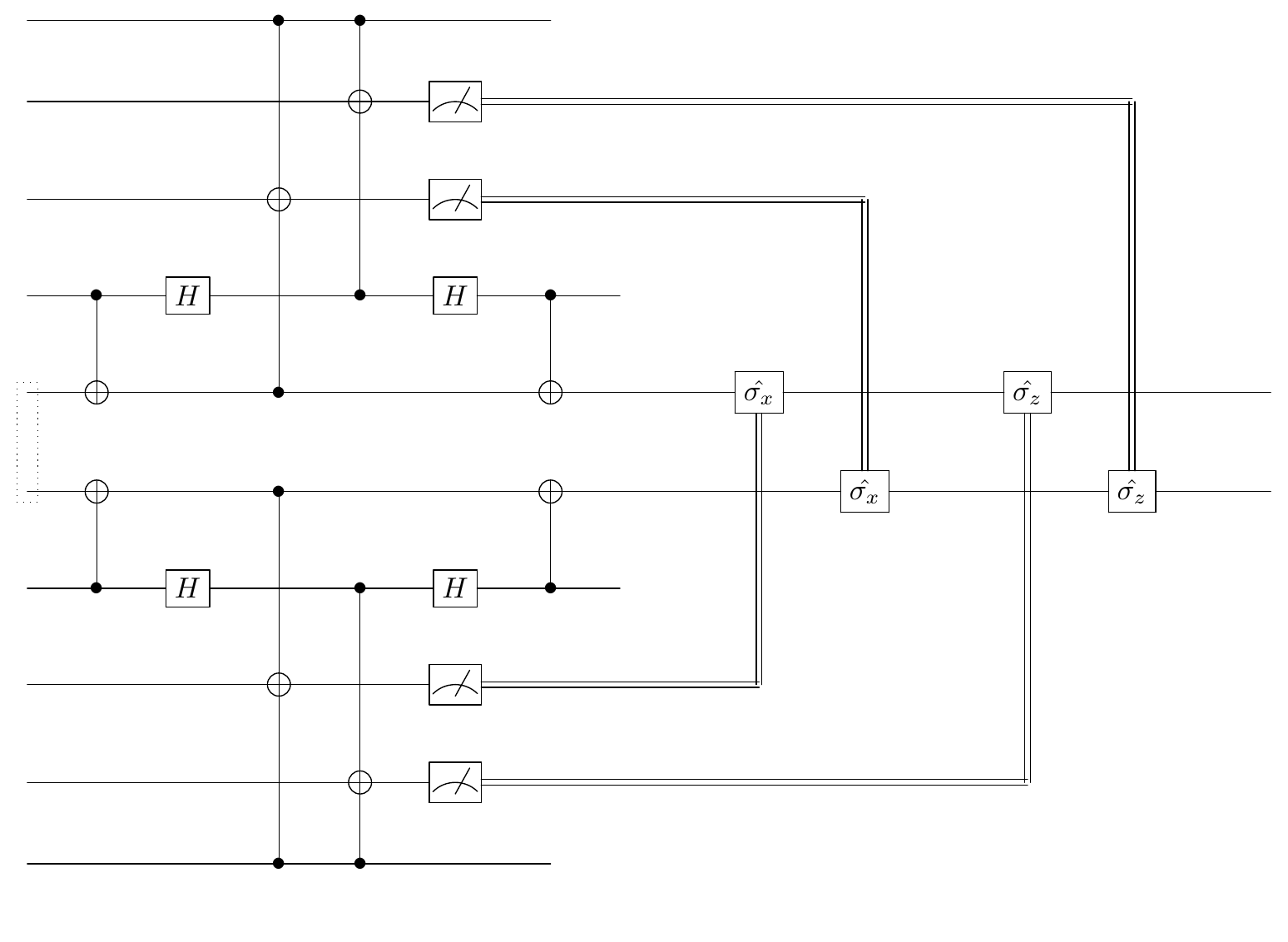}
\put(-330,220){$|\mathcal{T_A}\rangle$}
\put(-330,200){$|\mathcal{V}^{1}_A\rangle$}
\put(-330,177){$|\mathcal{V}^{2}_A\rangle$}
\put(-330,155){$|\mathcal{S_A}\rangle$}
\put(-340,120){$|\phi_{AB}\rangle$}
\put(-330,85){$|\mathcal{S_B}\rangle$}
\put(-330,57){$|\mathcal{V}^{1}_B\rangle$}
\put(-330,37){$|\mathcal{V}^{2}_B\rangle$}
\put(-330,17){$|\mathcal{T_B}\rangle$}
\put(-3,135){$\varrho^{(A)}_{out}$}
\put(-3,105){$\varrho^{(B)}_{out}$}
\caption{ Bidirectional quantum Teleportation circuit. $\bullet$ and $\oplus$ respectively indicate the control and target qubits  of $Cnot$ and $CCnot$ gates. Likewise, ${\hat{\sigma}_x}$, ${\hat{\sigma}_z}$, and $H$  respectively denote the Pauli's and Hadamard's gates. The dotted-line block indicates that the noise affects the circuit during the distribution of the state $\ket{\phi_{AB}}$ between the two partners. }
\label{FIGURE0}
 \end{center}
\end{figure}

 To implement the bidirectional scheme, Alice and Bob share the maximally entangled state:
\begin{eqnarray}\label{bell}
\ket{\phi_{AB}}=\frac{1}{\sqrt{2}}(\ket{00}+\ket{11})_{AB}.
\end{eqnarray}
By means of Pauli operators, we rewrite the state given in (Eq. \ref{bell}) as:
\begin{eqnarray}\label{Bell}
\hat{\mathcal{\varrho}}_{AB}=\frac{1}{4}(\hat{I}+\hat{\gamma}_1^{(A)}\hat{\delta}_1^{(B)}
-\hat{\gamma}_2^{(A)}\hat{\delta}_2^{(B)}+\hat{\gamma}_3^{(A)}\hat{\delta}_3^{(B)}).
\end{eqnarray}
Additionally, they perform projective measurements  and save their results on the qubits $\ket{\mathcal{V}^{(1)}_m}, \ket{\mathcal{V}^{(2)}_m}$ (for $m=A,B$), as shown  above in the quantum circuit (Fig. \ref{FIGURE0}). In fact, the storage qubits $\ket{\mathcal{V}^{(1)}_m}, \ket{\mathcal{V}^{(2)}_m}$ (for $m=A,B$) are initially prepared in the vacuum state: $\ket{\mathcal{V}^{(k)}_m}$=$|0\rangle$ (for $k=1,2$ and $m=A,B$). Furthermore, the two partners have all the elements to  perform the $BQT$ quantum circuit as shown in (Fig. \ref{FIGURE0}).

\subsection{Implementing the $BQT$ protocol}

They start the $BQT$ implementation by the state $\ket{\Delta_{s}}$:
\begin{equation}
 \ket{\Delta_{s}}=\ket{\mathcal{S_A} \mathcal{V}^{(1)}_A \mathcal{V}^{(2)}_A\mathcal{T_A}} \ket{\mathcal{S_B} \mathcal{V}^{(1)}_B \mathcal{V}^{(2)}_B\mathcal{T_B}} \otimes \ket{\phi_{AB}}.
\end{equation}
\begin{itemize}
	\item {  \it{\textbf{The first step}}}:\\ Alice and Bob perform two $Cnot$ gates where $\ket{\mathcal{S_A}}$ and $\ket{\mathcal{S_B}}$ are the control qubits and the Bell pairs act as the target qubits. 
\begin{eqnarray*}
\ket{\Delta_s}^{'}=Cnot_{2}( \ket{\mathcal{S_A} \mathcal{V}^{(1)}_A \mathcal{V}^{(2)}_A\mathcal{T_A}} \ket{\mathcal{S_B} \mathcal{V}^{(1)}_B \mathcal{V}^{(2)}_B\mathcal{T_B}} \otimes \ket{\phi_{AB}}),
\end{eqnarray*}
Afterward, they  apply  two Hadamard gates on the qubits $\ket{\mathcal{S_A}}$ and $\ket{\mathcal{S_B}}$ in the previous output results. The whole state of the system, therefore,  is given as:
\begin{eqnarray}
\ket{\Delta_s}^{''}=H_{2}(Cnot_{2}(\ket{\mathcal{S_A} \mathcal{V}^{(1)}_A \mathcal{V}^{(2)}_A\mathcal{T_A}} \ket{\mathcal{S_B} \mathcal{V}^{(1)}_B \mathcal{V}^{(2)}_B\mathcal{T_B}} \otimes \ket{\phi_{AB}} )).
\end{eqnarray}
\item {  \it{\textbf{The second step:}}}\\
Alice and Bob perform two $Ccnot$ gates. The trigger qubits $\ket{\mathcal{T_{A,B}}}$ and the Bell qubits $\ket{\phi_{AB}} $ are the control qubits and the qubits $\ket{ \mathcal{V}_{A,B}^{(2)}}$ are  the target qubits of the first $Ccnot$ gate. Moreover, the trigger qubits $\ket{\mathcal{T_{A,B}}}$ and the qubits $\ket{\mathcal{S_{A,B}}} $ are the control qubits and the qubits $\ket{ \mathcal{V}_{A,B}^{(1)}}$ are  the target qubits of the second $Ccnot$ gate.
\begin{eqnarray}
\ket{\Delta_s}^{'''}=Ccnot_2\Bigl(H_2\Bigr(Cnot_2(\ket{Q_A S^{(1)}_A S^{(2)}_AT_A} \ket{Q_B S^{(1)}_B S^{(2)}_BT_B}\otimes \ket{\phi_{AB}})\Bigl)\Bigl).
\end{eqnarray}
\item {  \it{\textbf{The third step:}}}\\  The two partners store their projective measurements outcomes  in the qubits $\ket{\mathcal{V}_{A,B}^{(1,2)}}$  and  $\ket{\mathcal{V}_{A,B}^{(1,2)}}$. These results  are sent via classical channels.  Based on Alice's measurements, Bob performs the adequate Pauli operator ($\hat{\sigma}_x ,\hat{\sigma}_y,\hat{ \sigma}_z$) on his Bell pair to reconstruct  the teleported state by Alice $\ket{\mathcal{S_A}}$.
\\

The above steps describe the teleportation process from Alice to Bob. However, in order to teleport the state from Bob to Alice, Bob may follow the same steps that Alice performs to teleport the state $\ket{\mathcal{S_B}}$.
\end{itemize}
Finally, the teleported states  are  given by \cite{5,d}:
\begin{eqnarray}\label{teleportedstates}
\varrho^{(A)}_{out}&=& \mathcal{M_B} \bar{ \mathcal{M_A}}\hat\varrho_{\mathcal{S}}^{(Bob)}+(1-{ \mathcal{M_B}}\bar{ \mathcal{M_A}})\hat{\varrho_{0}},~~ \bar{ \mathcal{M_A}}=1-{ \mathcal{M_A}},
\nonumber\\
\varrho^{(B)}_{out}&=&{ \mathcal{M_A}}\bar{ \mathcal{M_B}}\hat\varrho_{\mathcal{S}}^{(Alice)}+(1-{ \mathcal{M_A}}\bar{ \mathcal{M_B}})\hat{\varrho_{0}},~~ \bar{ \mathcal{M_B}}=1-{ \mathcal{M_B}}.
\end{eqnarray}
The probability of performing a sharp measurement $\mathcal{M}_i$ is:
\begin{equation}
\mathcal{M}_i=\operatorname{Tr}[\hat\varrho_{\mathcal{T}}^{(k)}\hat\varrho_{\mathcal{S}}^{(k)}],\quad\quad~~ k=\mathcal{A},\mathcal{B}.
\end{equation}
From Eqs. (\ref{teleportedstates}), one can note that Alice could reconstruct the teleported state $\hat\varrho_{\mathcal{S}}^{(Bob)}$ if Alice does not perform a sharp measurement and Bob does, then $ \mathcal{M_A}=0$ and $ \mathcal{M_B}=1$. Nevertheless, Bob gets the teleported state from Alice $\hat\varrho_{\mathcal{S}}^{(Alice)}$ if he does not apply a sharp measurement and  Alice does. If both partners use sharp measurements, they will obtain the following mixed states: $\varrho^{(A)}_{out}=\varrho^{(B)}_{out}=\varrho_0$.
\subsection{Entanglement and $BQT$}
The fidelity \cite{7a} is a figure of merit to assess the $BQT$ protocol. Indeed, the fidelity ($f$) is a closeness quantifier between the input and the output states of the teleportation protocol \cite{allati}. It is given by:
\begin{equation}
f^{ in \rightarrow out}= \langle \mathcal{S}^{in}| \varrho_{out} |\mathcal{S}^{in}\rangle,
\end{equation}
whereas $|\mathcal{S}^{in}\rangle$ is the input state and $\varrho_{out}$ is the output state. Since the essence of quantum teleportation is the transmission of all the unknown quantum states,  we use the following average amount:
\begin{equation}
\mathcal{F}_{avg}^{ in \rightarrow out}=\frac{1}{\Omega}\int_{0}^{\pi}d\theta_k\int_{0}^{2\pi}f^{ in \rightarrow out}(\theta_k,\phi_k)\sin\theta_k d\phi_k~ \quad k=\mathcal{A},\mathcal{B}.
\end{equation}
 $\Omega=4\pi$, is the solid angle.\\

Entanglement is a primordial ingredient for $BQT$. For the sake of measuring the entanglement in the quantum channel $\hat{\mathcal{\varrho}}_{AB}$ of the considered $BQT$ protocol, we use the negativity of the partial transposed density matrix \cite{Nega2}, $\mathcal{N}^{(PT)}$:

\begin{equation}
\mathcal{N}^{(PT)}= 2\sum^{3}_{n=0} \Lambda_{n},
\end{equation}
 $\Lambda_n$  are the negative eigenvalue of the partial transpose with respect to the qubit $i=A,B$, namely $\hat{\mathcal{\varrho}}_{AB}^{T_i}$. In addition, we will investigate quantum Fisher information $(QFI)$, which is more practical in a one- parameter teleportation scenario. Generally, a qubit state is given in the Bloch representation by:
\begin{equation}\label{ee}
\hat{\varrho}=\frac {1}{2}(\mathbf{I}+\overrightarrow{v}.\hat{\gamma}),
\end{equation}
where the vector $\overrightarrow{v}=(v_{x}, v_{y}, v_{z})^{T}$ indicates  the real Bloch vector and $\hat{\gamma}=(\hat{\gamma}_{x}, \hat{\gamma}_{y}, \hat{\gamma}_{z} ) $ stands for the Pauli operators. A simple expression of $QFI$, based on the Bloch presentation, is given as \cite{444}:
\begin{equation}\label{BlochQF}
\mathcal{J}_{\omega} = \left\{
\begin{array}{ll}
|\partial_{\omega} \overrightarrow{v}|^{2}+\frac{(\overrightarrow{v}.\partial_{\omega}\overrightarrow{v})^{2}}{1-|\overrightarrow{v}|^{2}}  & \mbox{if } |\overrightarrow{v}|<1 \\
|\partial_{\omega} \overrightarrow{v}|^{2} & \mbox{if }
|\overrightarrow{v}|=1,
\end{array}
\right.
\end{equation}
where: $\mathcal{J}_{\omega}$ is the quantum Fisher information with respect to the parameter $\omega$, with $\partial_{\omega}=\frac{\partial}{\partial_{\omega}}$. 
\\

The first line in  Eq. (\ref{BlochQF}), where the Bloch vector $|\overrightarrow{v}|<1 $, is for mixed states. Moreover,  for pure states with a Bloch vector, the quantum Fisher information is equal to unity ($ |\overrightarrow{v}|=1$),  and can be calculated as: $|\partial_{\omega} \overrightarrow{v}|^{2}$.

\section{Bidirectional teleportation via memory channels}
In what follows, we briefly present the notion of the partially correlated channel. Moreover, we   investigate the $BQT$ under typical correlated decoherence channels such as dephasing and amplitude damping channels.

\subsection{Correlated channel}
A quantum channel is a completely positive trace preserving map \cite{12} which describes the interaction of the quantum state $\hat{\varrho}$ with its noisy environment. The channel $\mathcal{X}$ maps from the input state $\hat{\varrho}$ to another state $\mathcal{X} (\hat{\varrho})$ which is given as \cite{1}:
\begin{equation}
\mathcal{X}(\hat{\varrho})=\sum_{i}\mathcal{K}_i \hat{\varrho} \mathcal{K}^{\dagger}_i.
\end{equation}
Where $\mathcal{K}_i=\sqrt{q_{i1,...,iN}}A_i$ are Kraus operators. $q_{i1,...,iN}$ are the probabilities of applying a sequence of  operations $A_i$ to a sequence of $N$ qubits passing through the channel are given above. However, in real scenarios, it is possible to have correlations between the successive actions of the channel. Indeed, $Macchiavello$ et $al.$ \cite{p1} proposed a model that describes a partially correlated channel.  For the two uses of the channel, the  Kraus operators  take the form \cite{p1}:
\begin{equation}
\mathcal{K}_{i,j}=\sqrt{q_i[(1-u)q_j+u\delta_{i,j}]}A_i \otimes A_j,
\end{equation}
The parameter $0\leq u \leq 1$  represents the memory strength of the noisy channel $\mathcal{X}$. In this model, the outcome state is obtained as:
\begin{equation}\label{0}
    \mathcal{X}(\hat{\varrho})= (1-u) \sum_{i,j=0} \mathcal{K}^{I}_{ij} \hat{\varrho}_i \mathcal{K}_{ij}^{I \dag} + u \sum_{k=0} \mathcal{K}^{c}_{kk} \hat{\varrho}_i \mathcal{K}_{kk}^{c \dag}.
\end{equation}
The Kraus operators $\mathcal{K}$ encodes the action of the channel $ \mathcal{X}$ on the state $\hat{\varrho}$. The superscripts $"I"$ and $"c"$ indicate independent and correlated, respectively.

\subsection{BQT via Dephasing noise }
A dephasing channel is used to characterize the loss of quantum information that occurs without the loss of energy \cite{PD1}. In other words, the relative phase between the energy eigenstates of the system is lost, which leads to the decay of the off-diagonal elements of the system's density matrix. Moreover, the Kraus operators for a single qubit are given as:
\begin{eqnarray}\label{pp}
\mathcal{K}^{u}_{00}&=&q_0\mathcal{\hat{I}} \otimes \mathcal{\hat{I}}; \quad \mathcal{K}^{u}_{01}=\sqrt{q_0 q_1}\mathcal{\hat{I}} \otimes \mathcal{\hat{\sigma}}_z; \quad \mathcal{K}^{u}_{10}=\sqrt{q_1 q_0}\mathcal{\hat{\sigma}}_z\otimes \mathcal{\hat{I}}
\nonumber\\
\mathcal{K}^{u}_{11}&=&q_1 \mathcal{\hat{\sigma}}_z \otimes \mathcal{\hat{\sigma}}_z;~\quad
 \mathcal{K}^{c}_{00}=\sqrt{q_0}\mathcal{\hat{I}} \otimes \mathcal{\hat{I}}; \quad \mathcal{K}^{c}_{11}=\sqrt{ q_1}\mathcal{\hat{\sigma}}_z \otimes \mathcal{\hat{\sigma}}_z,~\quad\quad\quad\quad\quad\quad\quad\quad\quad
\end{eqnarray}
 whereas $ q_0 = 1- \mathbf{p}_{D}$, $q_1 = \mathbf{p}_{D}$, and 
 $\mathbf{p}_{D}$ is the decoherence factor, $0 \leq \mathbf{p}_{D} \leq 1$. Supposing that the bidirectional teleportation quantum channel is shared via the dephasing noise channel  \cite{PD0}, the $BQT$ quantum channel can be straightly obtained by
substituting Eq.(\ref{pp}) and  the state Eq.(\ref{Bell}) into the Eq.(\ref{0}):
\begin{eqnarray}\label{Phase}
\hat{\mathcal{\varrho}}^{D}_{AB}=\frac{1}{2}
\begin{pmatrix}
1&0&0&1-4(1-u)(\mathbf{p}_{D}-\mathbf{p}_{D}^2)\\
0&0&0&0\\
0&0&0&0\\
1-4(1-u)(\mathbf{p}_{D}-\mathbf{p}_{D}^2)&0&0&1
\end{pmatrix}.
\end{eqnarray}
Now, to explore the $\mathbf{p}_D$ dependence on time, $Daffer$ $et$ $al.$ \cite{p2} presented a colored dephasing model, where $\mathbf{p}_{D}$ could explicitly depend on time. This model  describes the time evolution of a single qubit through the Hamiltonian $ \hat{H}(t)= \alpha \hbar \beta(t) \hat{\sigma}_z$, where $\alpha $ is a coin-flip random variable with possible values $\pm 1$. The random variable $\beta(t)$ has a Poisson distribution with a mean equal to $\frac{t}{2\tau}$. The decoherence factor can be given as\cite{p2}:
\begin{equation}
\mathbf{p}(t)_{D}=\frac{1-\gamma(t)}{2},
\end{equation}
 where
\begin{equation}
\gamma(t)=e^{-\nu}\bigr(\cosh(U\nu)+\frac{1}{U}\sinh(U\nu)\bigl),
\end{equation}
with $U=\sqrt{1-16\tau^2}$, and $\nu=\frac{t}{2\tau}$ is a dimensionless time. The parameter $\tau$ controls the degree of non-Markovianity of the dephasing process, where $\tau < 0.25$ corresponds to the Markovian regime. However,  $\tau > 0.25$ corresponds to the non-Markovian regime. As a result, the amount of survival quantum entanglement in the quantum channel Eq.(\ref{Phase}) is as follows:
\begin{equation}
\mathcal{N}^{(PT)}_{D}= 1-4(1-u)(\mathbf{p}_{D}-\mathbf{p}_{D}^2).
\end{equation}
Alice and Bob perform the $BQT$ protocol by means of the noisy quantum channel Eq.(\ref{Phase}). They can get the teleported states:
\begin{equation}
\hat{\varrho}_{A}^{D}=\frac{1}{2}\Bigl(\mathbf{I}+\sum_{n=1,2,3}\alpha^{(D)}_n\hat{\gamma_n}\Bigr), ~ \hat{\varrho}_{B}^{D}=\frac{1}{2}\Bigl(\mathbf{I}+\sum_{n=1,2,3}\beta^{(D)}_n \hat{\delta_n}\Bigr),
\end{equation}
with different elements
\begin{eqnarray}\label{10}
\alpha_{1}^{(D)}&=&{ \mathcal{M_A}} \bar{ \mathcal{M_B}} \sqrt{\mathcal{N}^{(PT)}_{D}} \cos(\varphi_A) \sin(\vartheta_A),\quad
\alpha_{2}^{(D)}= -{ \mathcal{M_A}} \bar{ \mathcal{M_B}} \sqrt{\mathcal{N}^{(PT)}_{D}} \sin(\varphi_A) \sin(\vartheta_A),\quad
\alpha_{3}^{(D)}={ \mathcal{M_A}} \bar{ \mathcal{M_B}} \cos(\vartheta_A).
\nonumber\\
\beta_{1}^{(D)}&=&{ \mathcal{M_B}} \bar{ \mathcal{M_A}} \sqrt{\mathcal{N}^{(PT)}_{D}} \cos(\varphi_B) \sin(\vartheta_B),\quad
\beta_{2}^{(D)}=-{ \mathcal{M_B}} \bar{ \mathcal{M_A}} \sqrt{\mathcal{N}^{(PT)}_{D}} \cos(\varphi_B) \sin(\vartheta_B),\quad
\beta_{3}^{(D)}={ \mathcal{M_B}} \bar{ \mathcal{M_A}}  \cos(\vartheta_B).\quad\quad\quad\quad\quad\quad
\end{eqnarray}
The teleportation fidelities  are given as,
\begin{eqnarray}
    f_{D}^{A \rightarrow B} &=&-\frac{ { \mathcal{M_A}} \bar{ \mathcal{M_B}}}{2} \bigl(1-\sqrt{\mathcal{N}^{(PT)}_{D}})\sin(\vartheta_A)^2 + \frac{1+{ \mathcal{M_A}} \bar{ \mathcal{M_B}}}{2}, \nonumber \\
       f^{B \rightarrow A}_{D} &=& -\frac{{ \mathcal{M_B}} \bar{ \mathcal{M_A}}}{2} \bigl(1-\sqrt{\mathcal{N}^{(PT)}_{D}})\sin(\vartheta_B)^2 + \frac{1+{ \mathcal{M_B}} \bar{ \mathcal{M_A}}}{2}.
\end{eqnarray}

By using the expressions in Eqs.(\ref{10}, \ref{BlochQF}) and after a straightforward calculation, we get the explicit expression of the quantum Fisher information of the parameter $\vartheta_i$, $\mathcal{J}_{\vartheta_i}$, with $i=A,B$:
\begin{equation}
\mathcal{J}_{\theta_A} = 
r+\frac{w^2}{1-|\overrightarrow{a}|^{2}}, 
\end{equation}
where:
\begin{equation}
r=\bar {\mathcal{M}}_B^2 \Bigr(\mathcal{N}^{(PT)}_{D} (\partial_{\vartheta_A}{\mathcal{M}}_A \sin(\vartheta_A)+{\mathcal{M}}_A \cos(\vartheta_A))^2 +((\partial_{\vartheta_A}{\mathcal{M}}_A \cos(\vartheta_A)-{\mathcal{M}}_A \sin(\vartheta_A)))^2 \Bigl);
\end{equation}
and
\begin{equation}
w={\mathcal{M}}_A \bar {\mathcal{M}}_B^2 \Bigr(\mathcal{N}^{(PT)}_{D} \sin(\vartheta_A)(\partial_{\vartheta_A}{\mathcal{M}}_A \sin(\vartheta_A)+ {\mathcal{M}}_A \cos(\vartheta_A))+\cos(\vartheta_A) (\partial_{\vartheta_A}{\mathcal{M}}_A \cos(\vartheta_A)^2 - {\mathcal{M}}_A\cos(\vartheta_A)\Bigl).
\end{equation}
\\  
Yet, the $QFI$ of the parameter $\vartheta_B$ is written as:
 \begin{equation}
\mathcal{J}_{\vartheta_B} = 
\mathcal{V}+\frac{\mathcal{W}^2}{1-|\overrightarrow{b}|^{2}}, 
\end{equation}
where:
\begin{equation}
\mathcal{V}=\bar {\mathcal{M}}^2 \Bigr(\mathcal{N}^{(PT)}_{D} (\partial_{\vartheta_B}{\mathcal{M}}_B \sin(\vartheta_B)+{\mathcal{M}}_B \cos(\vartheta_B))^2 +((\partial_{\vartheta_B}{\mathcal{M}}_B \cos(\vartheta_B)-{\mathcal{M}}_B \sin(\vartheta_B)))^2 \Bigl);
\end{equation}
and
\begin{equation}
\mathcal{W}=p_A \bar {\mathcal{M}}_B^2 \Bigr((\partial_{\vartheta_B}{\mathcal{M}}_A \sin(\vartheta_A)^2 + {\mathcal{M}}_A \cos(\vartheta_A)\sin(\vartheta_A))+(\mathcal{N}^{(PT)}_{D})^2 (\partial_{\vartheta_B}{\mathcal{M}}_A \cos(\vartheta_A)^2 - {\mathcal{M}}_A\cos(\vartheta_A)\sin(\vartheta_A)) \Bigl)
\\
\end{equation}
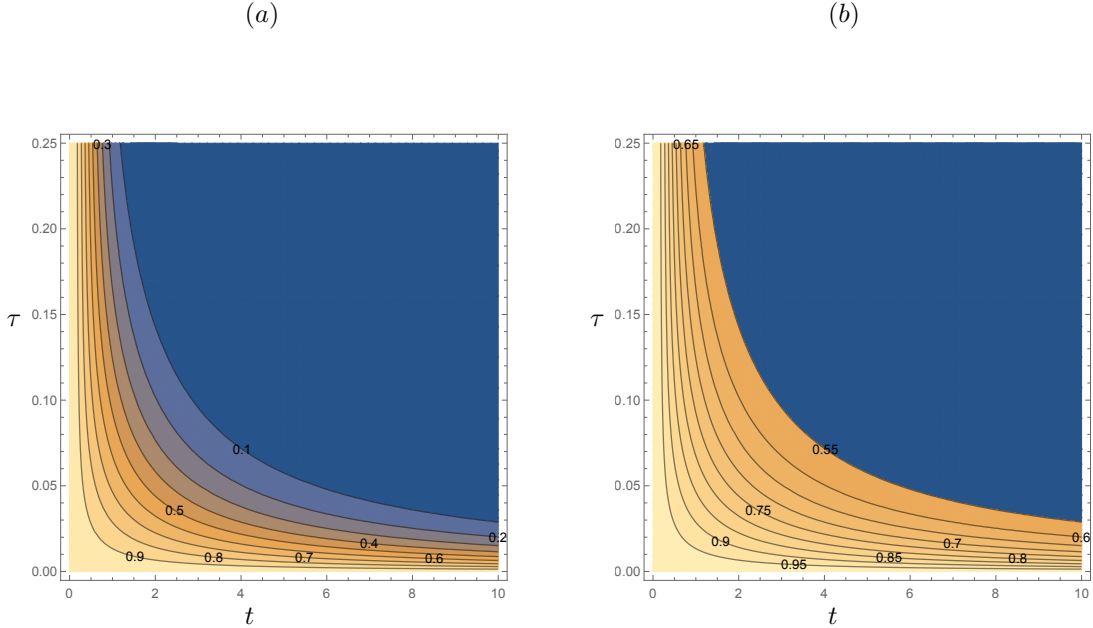
\begin{figure}[!htb]
\begin{center}

\includegraphics[scale=.5]{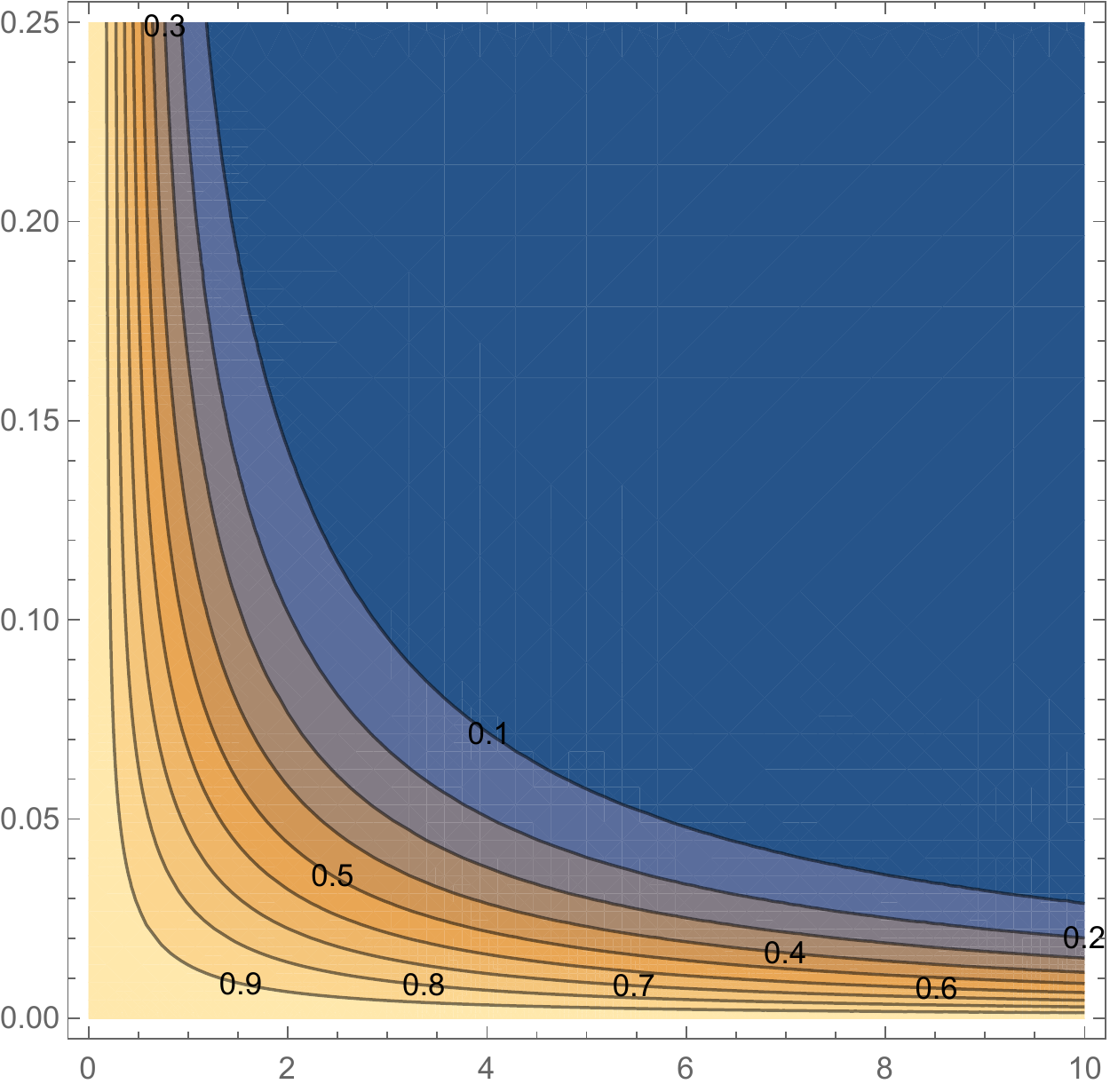}
\put(-190,105){$\tau$}
                      \put(-100,220){$(a)$}
                         \put(-100,-8){$t$}\quad\quad\quad\quad
\includegraphics[scale=.5]{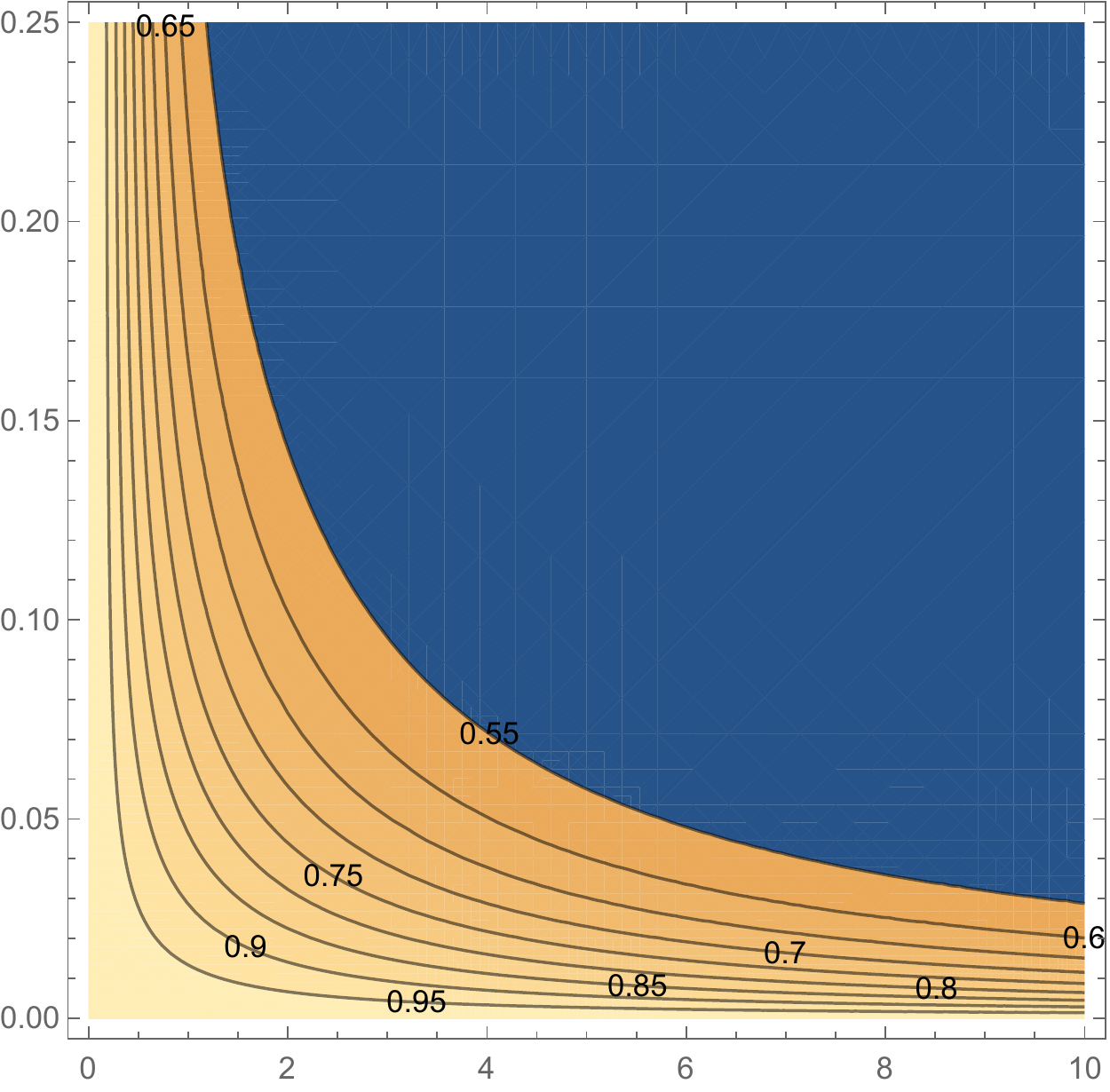}
\put(-190,105){$\tau$}
                      \put(-100,220){$(b)$}
                         \put(-100,-8){$t$}
\caption{Negativity of the partial transpose $\mathcal{N}^{(PT)}_{D}$, for the dephasing channel in the Markovian regime with $\tau<0.25$, versus  time  $t$. (a): $u=0$ and (b): $u=0.5$.
}
\label{FIGURED0}
 \end{center}
\end{figure}

The dynamics of the entanglement of independent ($u = 0$) and correlated ($u = 0.5$) dephasing channels in the Markovian regime, namely for $\tau<\frac{1}{4}$, are depicted in Fig.\ref{FIGURED0}. In fact, Fig.\ref{FIGURED0}.(a) shows that the negativity of the partial transpose decreases gradually over time and by increasing the Markovianity strength, $\tau$. Besides, in Fig.\ref{FIGURED0}.(b), we consider that the dephasing channel is partially correlated, where the correlation strength is $u=0.5$. By inspection of Fig.\ref{FIGURED0}.(b), it is obvious that, the negativity of the partial transpose is significantly increased thanks to the classical correlations in the dephasing channel. Thus, it is interesting to investigate the effects of the classical correlations in the Markovian regime on the entanglement, the average fidelities, and the quantum Fisher information. The parameter $tau=0.1$ is now set,
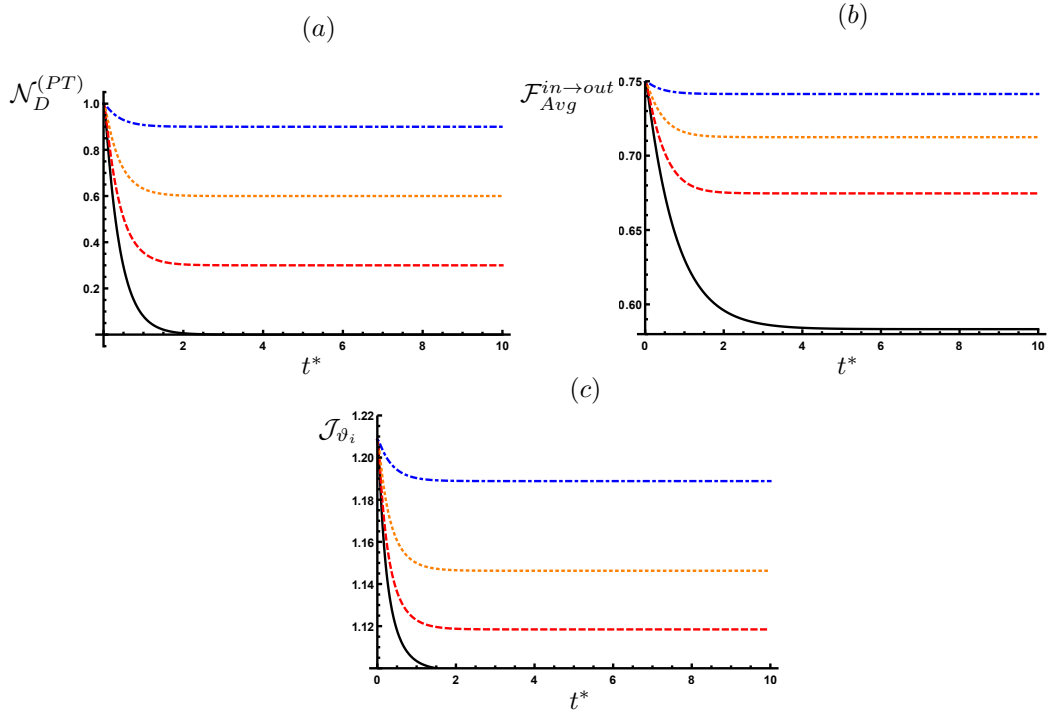
\begin{figure}[!htb]
\begin{center}
\includegraphics[scale=.45]{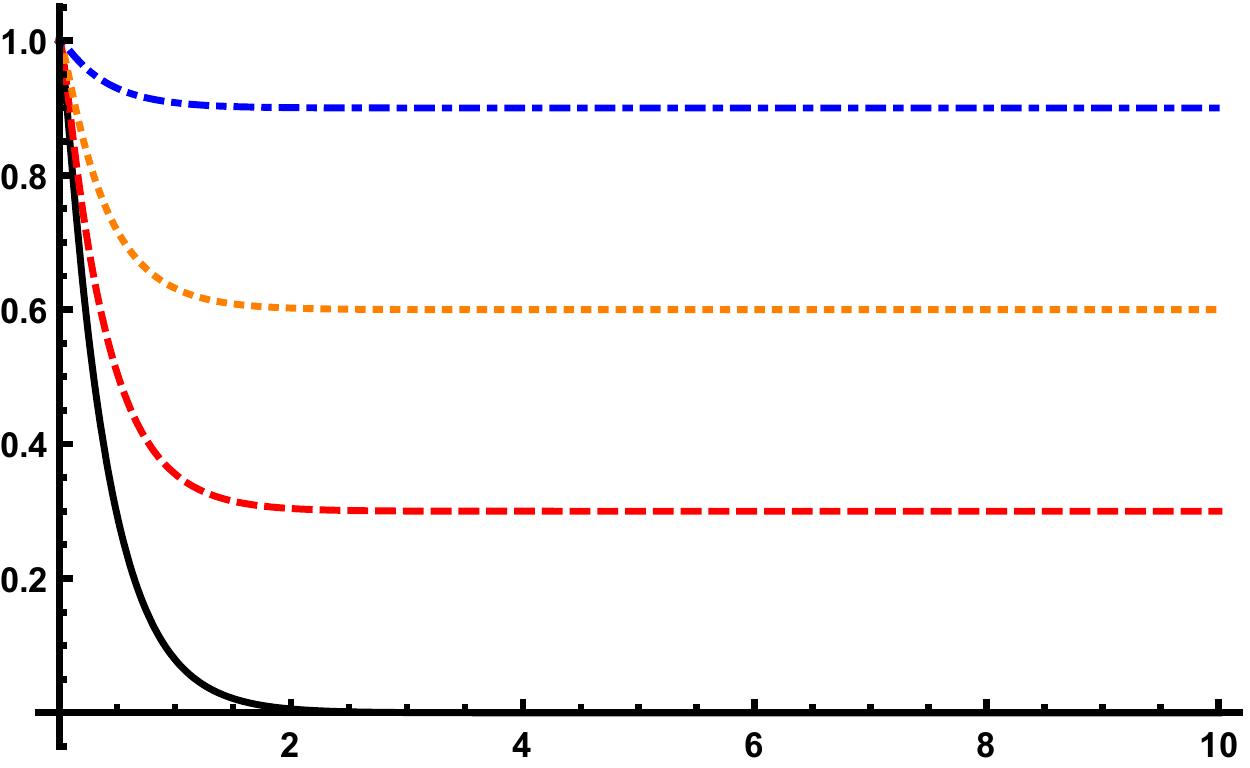}
\put(-190,95){$\mathcal{N}^{(PT)}_{D}$}
                      \put(-80,120){$(a)$}
                         \put(-80,-8){$t^{*}$}\quad\quad\quad\quad
\includegraphics[scale=.45]{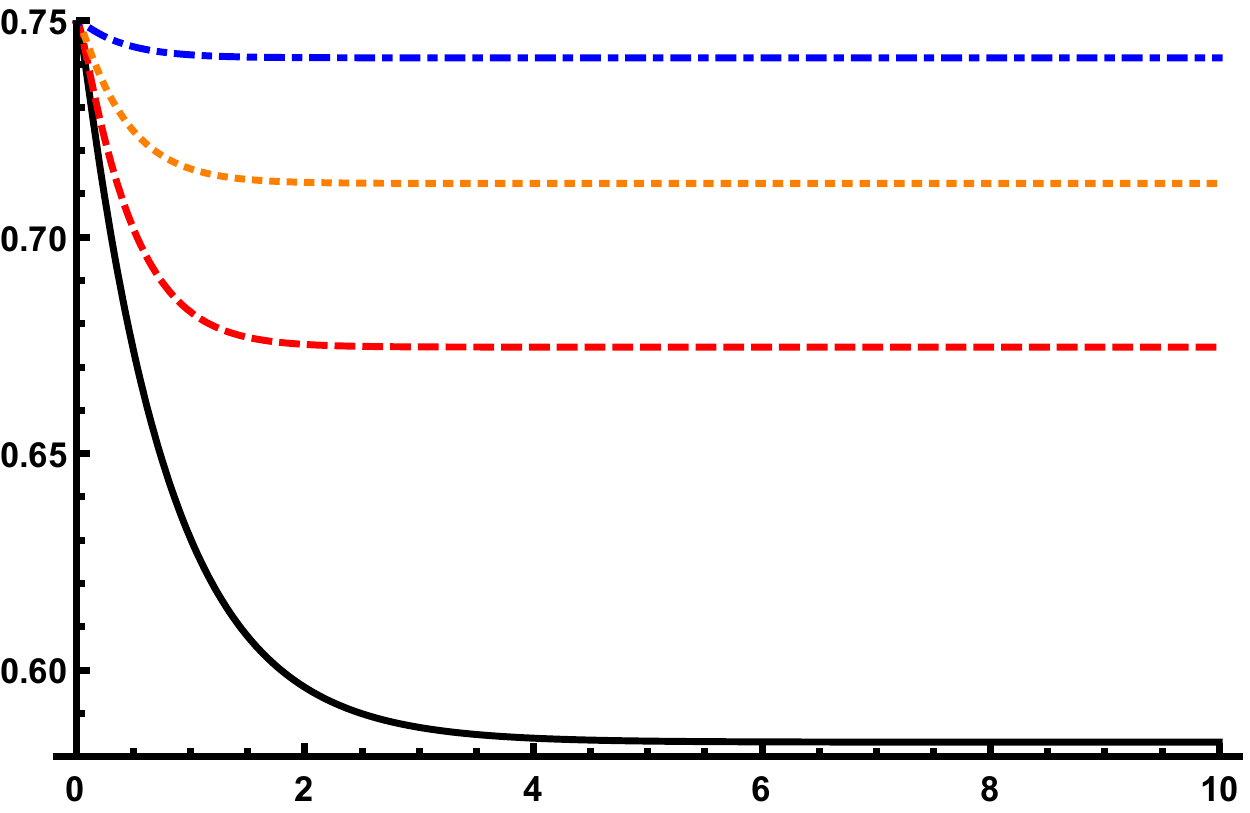}
\put(-200,95){$\mathcal{F}^{in\rightarrow out}_{Avg}$}
                      \put(-80,125){$(b)$}
                         \put(-80,-8){$t^{*}$}
\\
\includegraphics[scale=.45]{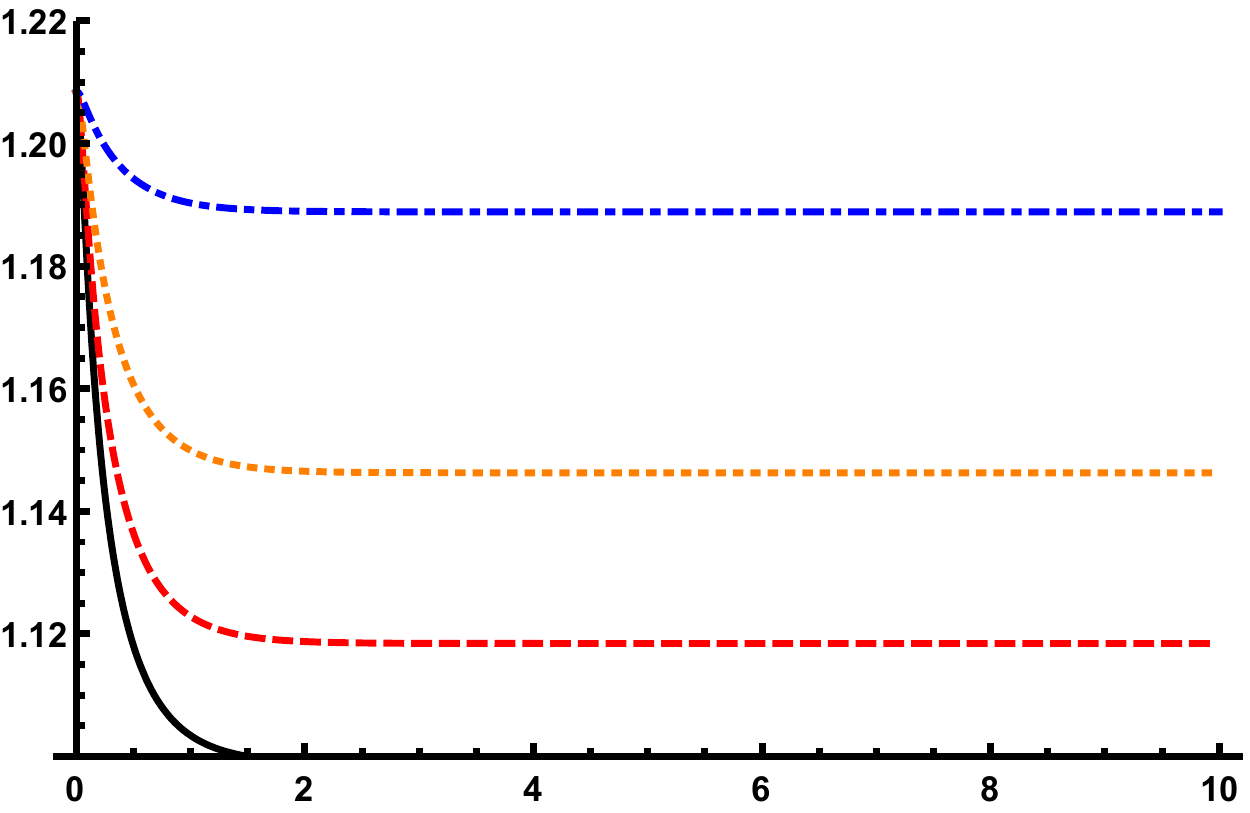}
\put(-175,95){$\mathcal{J}_{\vartheta_i}$}
                      \put(-80,110){$(c)$}
                         \put(-80,-8){$t^{*}$}
\caption{$\mathcal{N}^{(PT)}_{D}$, $\mathcal{F}^{in\rightarrow out}_{Avg}$ and $\mathcal{J}_{\vartheta_i}$ for $i=A,B$, versus the re-scaled  time  $t^{*}=\frac{t}{\pi}$ for the dephasing channel in the Markovian regime with $\tau=0.1$. The solid black, Red dashed, Orange dotted and blue dashdotted correspond to $u=0, 0.3, 0.6$, and $0.9$, respectively.
}
\label{FIGURED1}
 \end{center}
\end{figure}
in Fig.\ref{FIGURED1}, we show the negativity of the partial transpose, the average fidelity from the sender $"in"$ to the receiver $"out"$, and the quantum Fisher information of the parameter $\vartheta_i$ where $i=A$ or $B$, namely $\mathcal{N}^{(PT)}_{D}$, $\mathcal{F}^{in\rightarrow out}_{Avg}$ and $\mathcal{J}_{\vartheta_i}$  versus time in the Markovian regime $(\tau=0.1)$, where the quantum channel of the $BQT$ passes through a colored dephasing channel. In this case, we consider $\tau =0.1$, where $U \approx 1-\tau$. Therefore, we find
\begin{eqnarray}
\mathbf{p}(t)_{D}=(1-\frac{e^{\kappa t} (1+\kappa) - e^{\kappa}(1-\kappa)}{2\kappa})/2,
\end{eqnarray}
whereas: $\kappa=1-\tau$. It is obvious from Fig.\ref{FIGURED1} that the three quantities are decreasing monotonically with the flow of time $t$. However, by increasing the correlation factor $u$, the three quantities could be increased  and reach their asymptotic values, $\mathcal{N}^{(PT)}_{D} \approx u$, $\mathcal{F}^{in\rightarrow out}_{Avg} \approx \frac{u}{n}+0.5$ (for $n=1,2,3,4$) and $\mathcal{J}_{\vartheta_i}=0.8+u$ ($i=A,B$). Also, the behavior of the three quantities depends  on the parameter $\tau$. Displaying the effect of a larger value of $\tau$, in particular for $\tau>\frac{1}{4}$, on these quantities is, therefore, worth it. 
\\

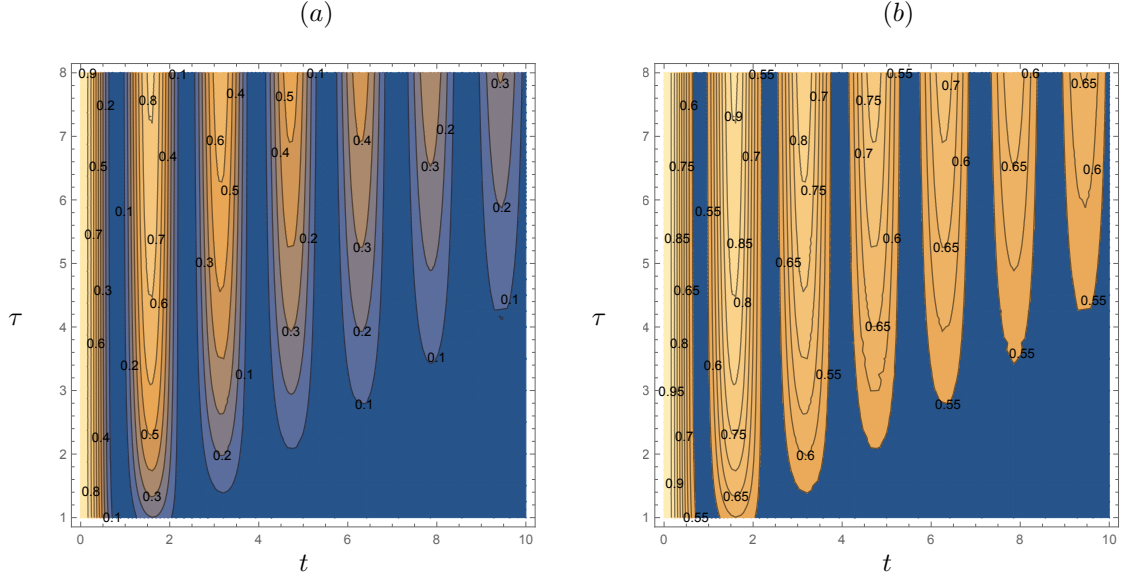
\begin{figure}
\begin{center}
\includegraphics[scale=.5]{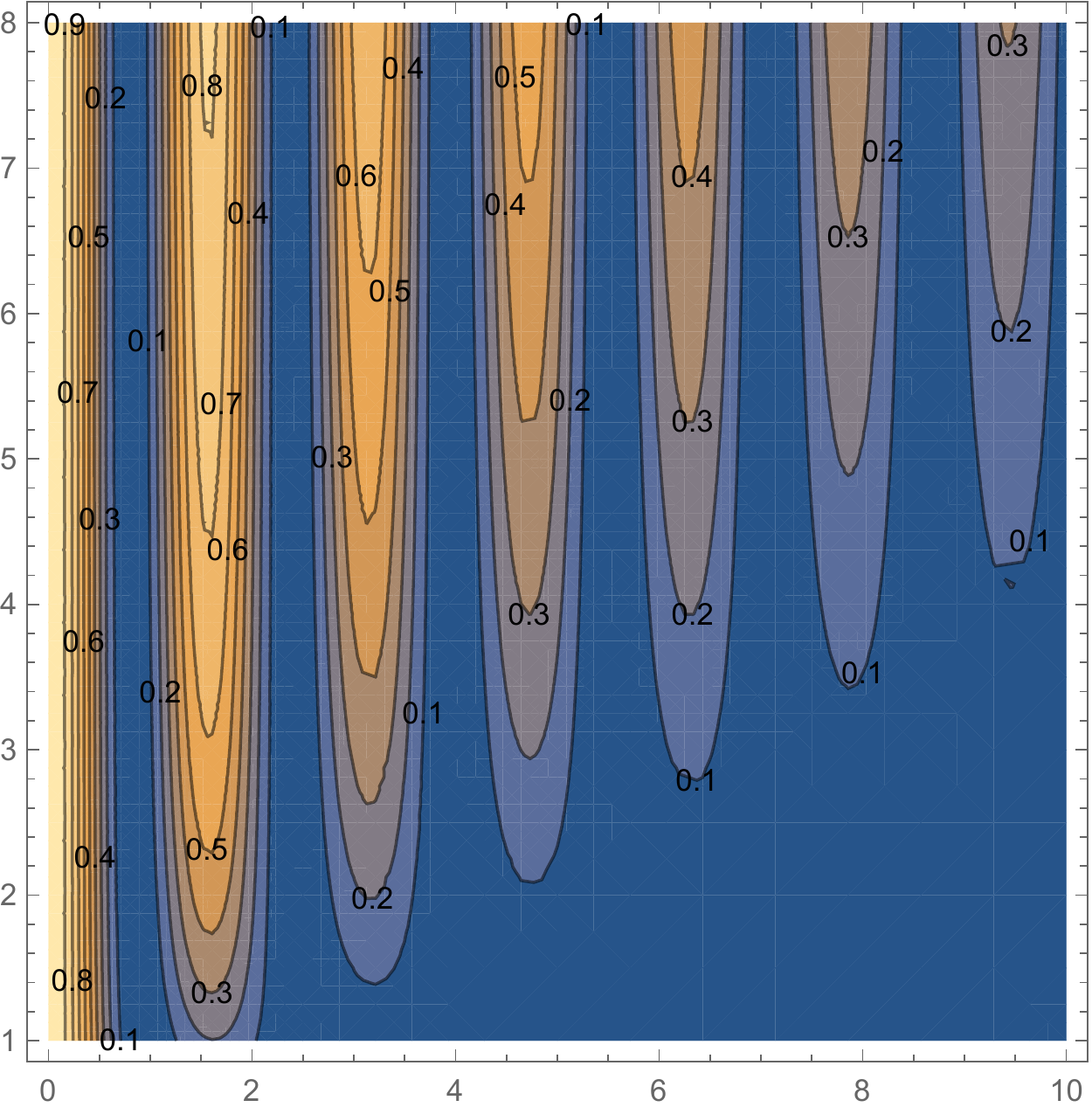}
\put(-200,85){$\tau$}
                      \put(-90,200){$(a)$}
                         \put(-90,-8){$t$}\quad\quad\quad\quad
\includegraphics[scale=.5]{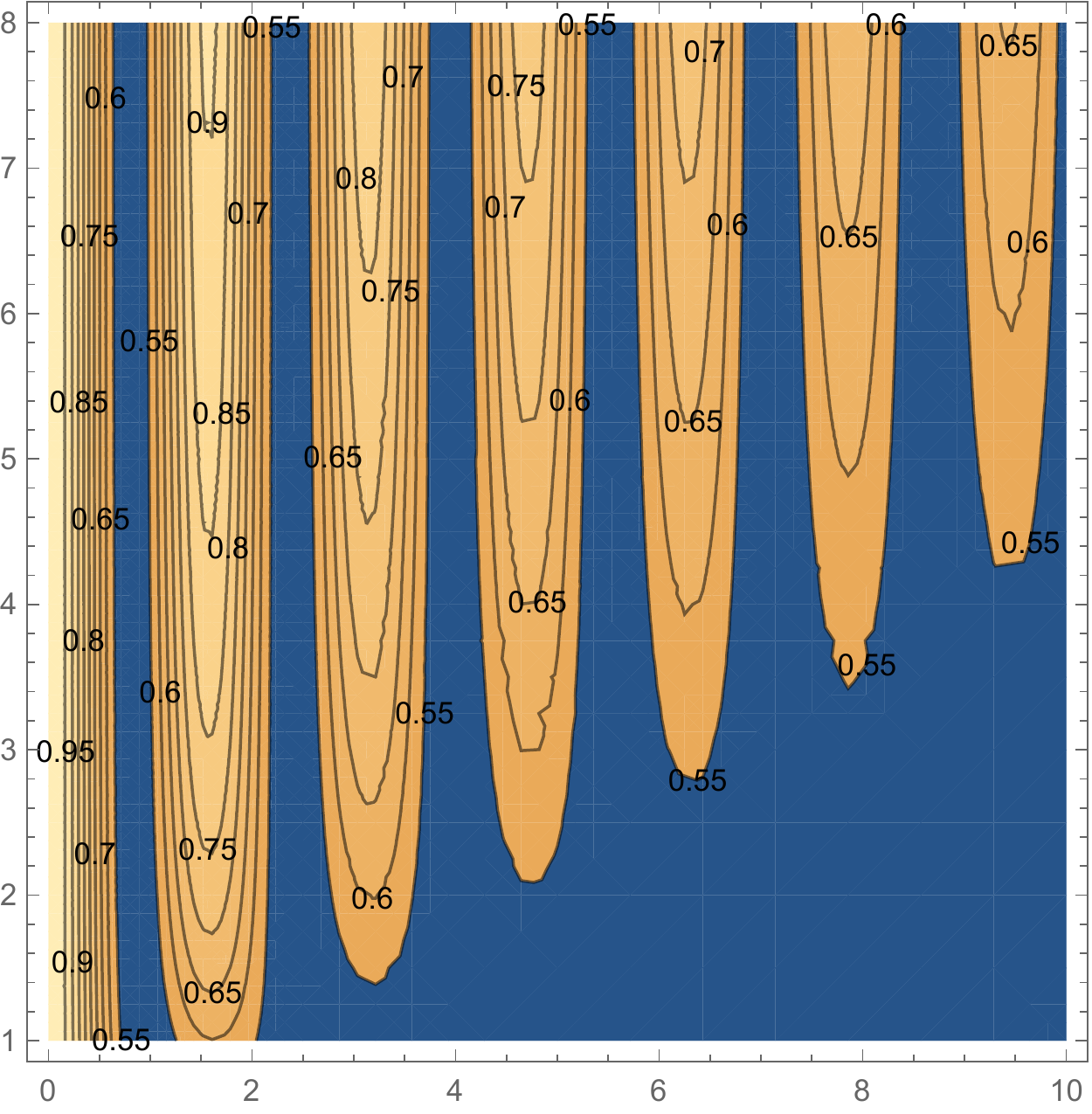}
\put(-200,85){$\tau$}
                      \put(-90,200){$(b)$}
                         \put(-90,-8){$t$}
\caption{The negativity of the partial transpose $\mathcal{N}^{(PT)}_{D}$, for the dephasing channel in the non-Markovian regime with $\tau>0.25$, versus  time  $t$. (a): $u=0$ and (b): $u=0.5$.}
\label{FIGURED2a}
 \end{center}
\end{figure}
 Fig.\ref{FIGURED2a} shows the periodic behavior of the  negativity of the partial transpose for the independent ($u=0$) and the correlated ($u=0.5$) dephasing channels in the non-Markovian regime, namely for $\tau>\frac{1}{4}$. It is apparent from Fig.\ref{FIGURED2a}.(a) that,  even in the non-Markovian regime, the contours of the entanglement decrease as time goes by. Besides, the amount of entanglement in the quantum channel increases as the value of $\tau$ increases. Furthermore, the dynamic of the entanglement for the correlated dephasing channel, where $u=0.5$, is shown in Fig.\ref{FIGURED2a}.(b). Examination Fig.\ref{FIGURED2a}.(b), it is clear that by considering the correlations in the dephasing channel,  one can enhance the strength of the negativity of the partial transpose $\mathcal{N}^{(PT)}_{D}$.
\\

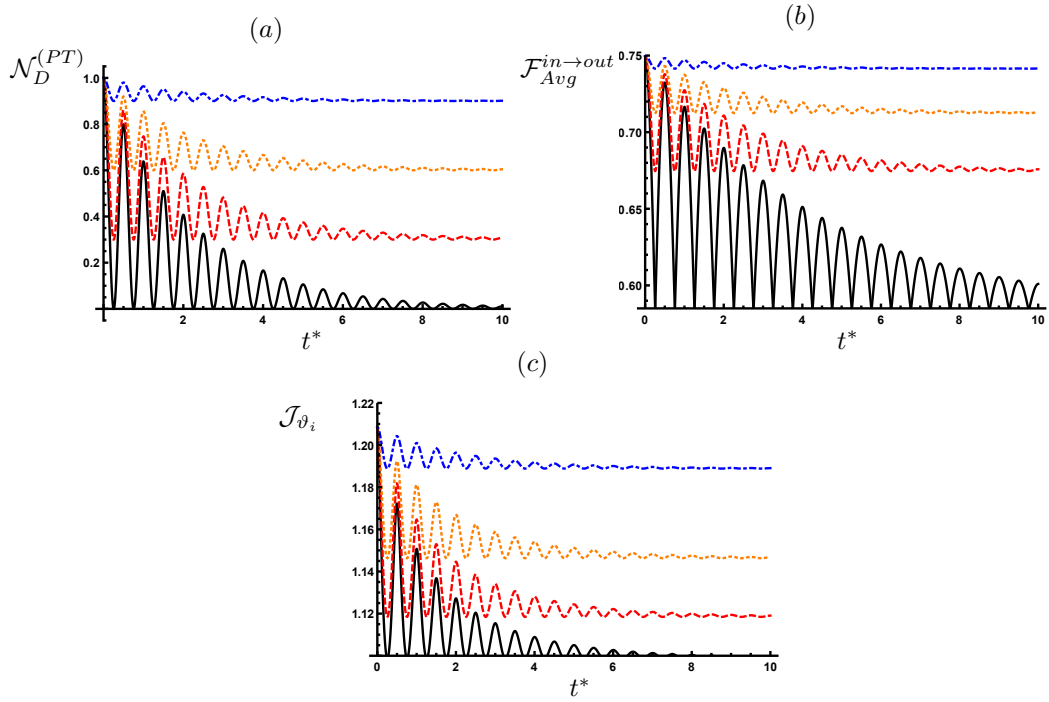
\begin{figure}
\begin{center}
\includegraphics[scale=.45]{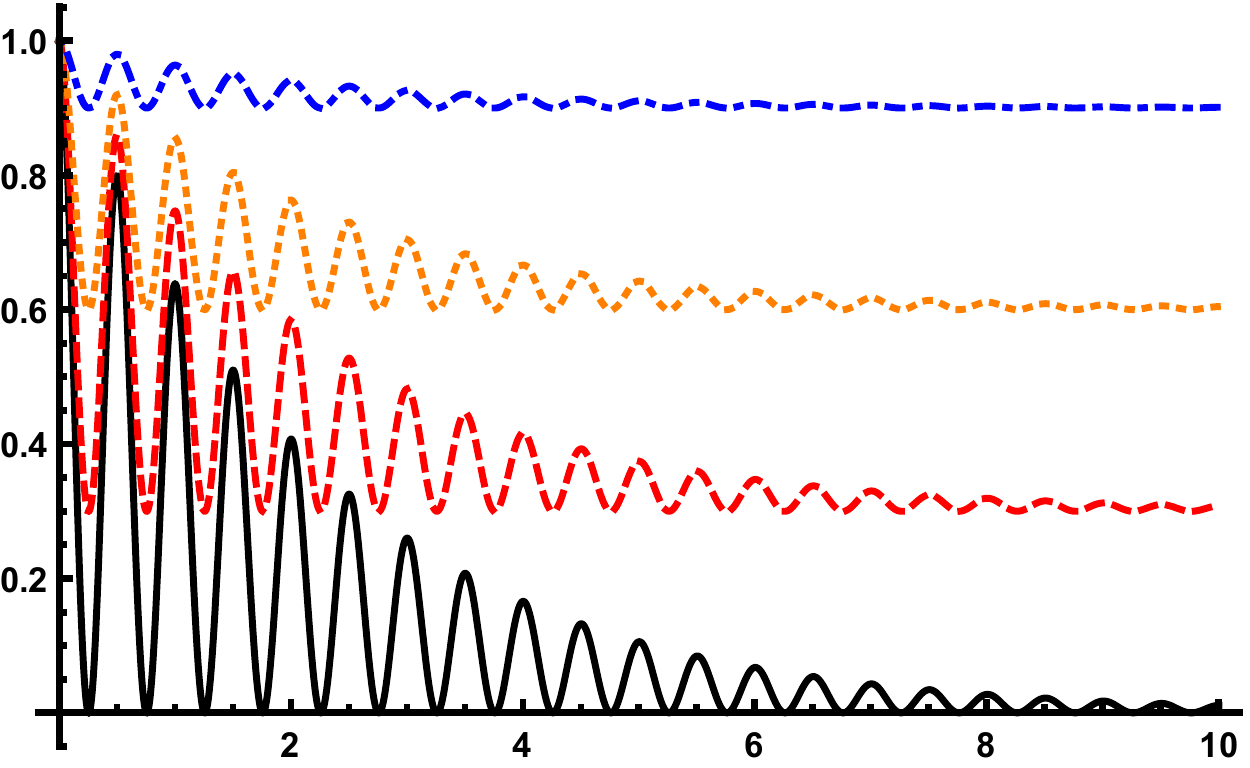}
\put(-190,95){$\mathcal{N}^{(PT)}_{D}$}
                      \put(-100,110){$(a)$}
                         \put(-80,-8){$t^{*}$}
\quad\quad\quad\quad
\includegraphics[scale=.45]{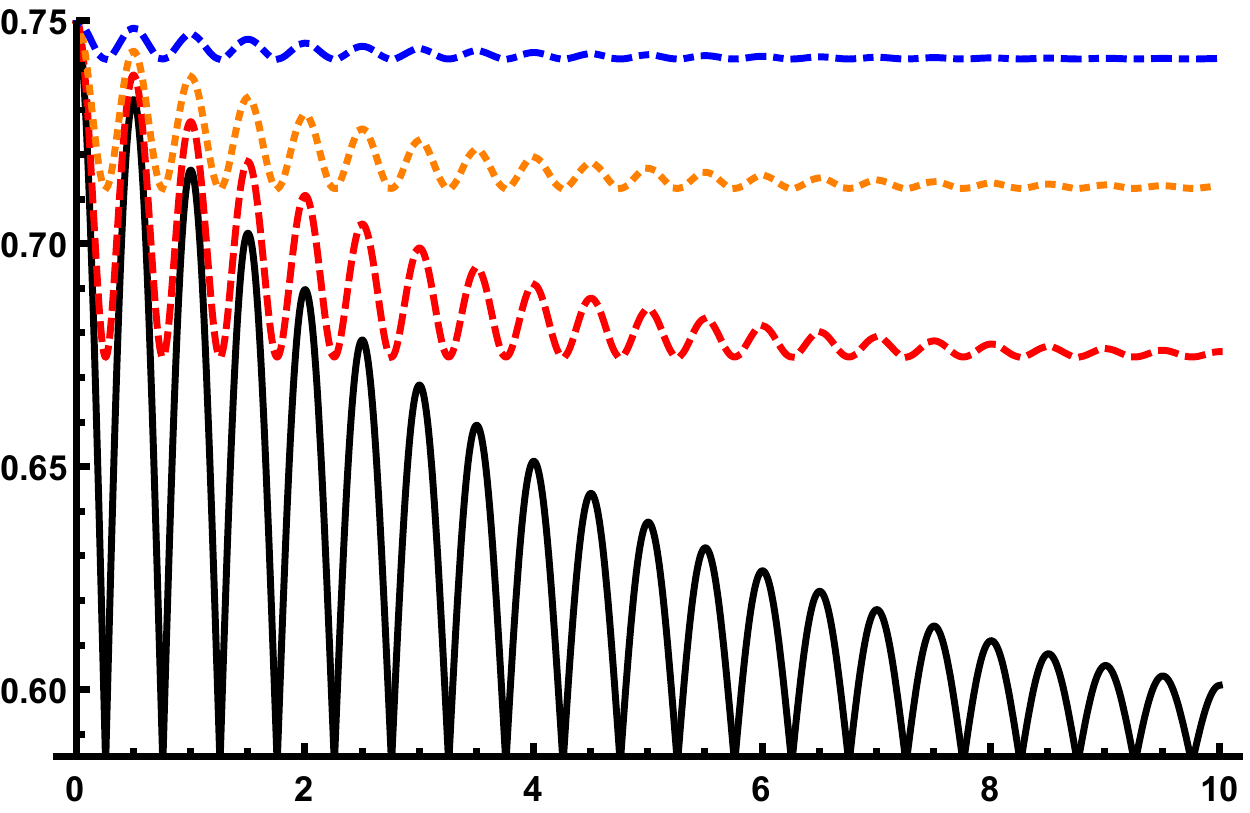}
\put(-200,95){$\mathcal{F}^{in\rightarrow out}_{Avg}$}
                      \put(-100,115){$(b)$}
                         \put(-80,-8){$t^{*}$}\\
\includegraphics[scale=.45]{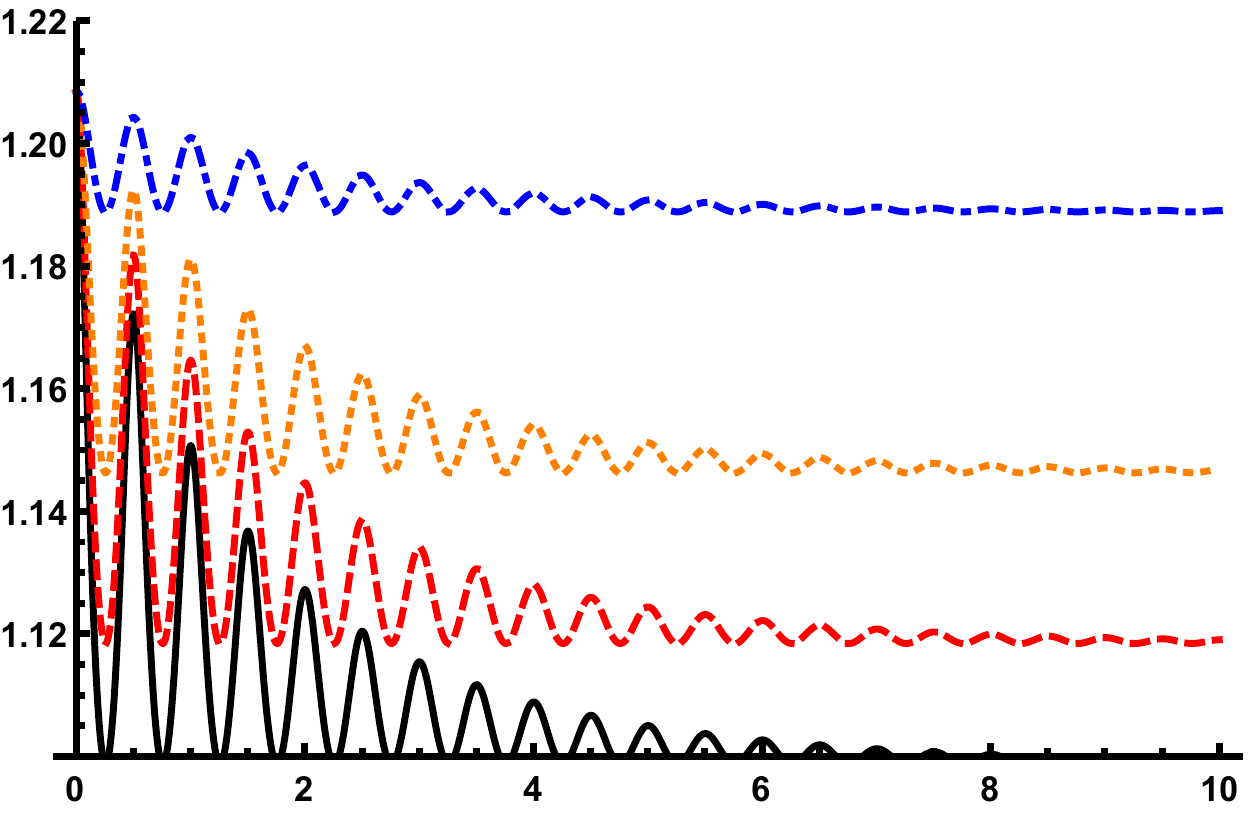}
\put(-190,95){$\mathcal{J}_{\vartheta_i}$}
                      \put(-100,115){$(c)$}
                         \put(-80,-8){$t^{*}$}
\caption{$\mathcal{N}^{(PT)}_{D}$, $\mathcal{F}^{in\rightarrow out}_{Avg}$ and $\mathcal{J}_{\vartheta_i}$ for $i=A,B$, versus the re-scaled  time  $t^{*}=\frac{t}{\pi}$ for the dephasing channel in the non- Markovian regime with $\tau=7$. The black, Red-dashed, Orange-dotted and blue dash-dotted curves correspond to $u=0, 0.3, 0.6$, and $0.9$, respectively.}
\label{FIGURED2}
 \end{center}
\end{figure}
The behavior of the negativity, average fidelity, and quantum Fisher information of the parameter $\vartheta$, namely $\mathcal{N}^{(PT)}_{D}$, $\mathcal{F}^{in\rightarrow out}_{Avg}$ and $\mathcal{J}_{\vartheta_i}$, with $i=A,B$ versus time in the non-Markovian regime of the colored dephasing channel, where $\tau > 0.25$, is shown in Fig.\ref{FIGURED2}. It is clear from these figures that the behavior of the three quantities shows damped oscillations. If $\tau>0.25$,  we get $|U| \approx 4\tau$, and $\mathbf{p}(t)_{D}=(1-e^{-\frac{t}{2\tau}\cos(2t)})/2$. it is clear from the figure that $T=0.5\pi$ which is the oscillation period. At the instants $t=n(T)$ with $n\in \mathbb{N}$, the negativity, the average fidelity and the QFI reach their maximum values. These maximum values decrease with time. However, at instants $t=m\frac{T}{2}$, where $m \in \mathbb{N}^{*}$, the three quantities namely $\mathcal{N}^{(PT)}_{D}$, $\mathcal{F}^{in\rightarrow out}_{Avg}$ and $\mathcal{J}_{\vartheta_i}$ decrease to their minimum values, $u$, $\frac{u}{n}+0.5$, 0.8+$u$, respectively. Hence, the exploitation of the combined effects of memory: the classical and temporal correlations between successive actions of the dephasing channel, could further enhance the amount of entanglement remaining in the quantum channel, which is distributed by a dephasing channel, and the teleportation of information bidirectionally. 
\newpage
\subsection{Amplitude damping channel}
The amplitude damping channel describes the release of the qubit's energy to its surrounding environment\cite{AD1}. It also describes the dissipating nature of the system-environment interaction \cite{AD2}. The Kraus operators that  represent the amplitude damping channel  are given as \cite{AD3}:
\begin{eqnarray}
\mathcal{K}^{u}_{00}=\mathcal{K}_0 \otimes \mathcal{K}_0; \quad \mathcal{K}^{u}_{01}=\mathcal{K}_0 \otimes \mathcal{K}_1; \quad \mathcal{K}^{u}_{10}=\mathcal{K}_1 \otimes \mathcal{K}_0;\quad \mathcal{K}^{u}_{11}=\mathcal{K}_1 \otimes \mathcal{K}_1,
\end{eqnarray}
where
\begin{eqnarray}\label{ampli}
\mathcal{K}_0 = 
\begin{pmatrix}
1&0\\
0&\sqrt{1-\mathbf{p}_{A}}
\end{pmatrix}; \quad
\mathcal{K}_1 = 
\begin{pmatrix}
0&\sqrt{\mathbf{p}_{A}}\\
0&0
\end{pmatrix},
\end{eqnarray}
with $\mathbf{p}_{A}$ is the amplitude damping probability. The Kraus operators for the correlated amplitude damping channel could be given as \cite{AD4}:
\begin{equation}\label{aa}
\mathcal{K}^{c}_{00}=\ket{00}\bra{00}+\ket{01}\bra{01}+\ket{10}\bra{10}+\sqrt{1-\mathbf{p}_{A}}\ket{11}\bra{11}, \quad \mathcal{K}_{11}^{c}=\sqrt{\mathbf{p}_{A}}\ket{00}\bra{11}.
\end{equation}
By replacing the Kraus operators  Eqs.(\ref{ampli}) and (\ref{aa}) in the channel Eq.(\ref{0}), we get the quantum channel of the BQT:
\begin{eqnarray}\label{Ad}
\hat{\mathcal{\varrho}}^{Ad}_{AB}=\frac{1}{2}
\begin{pmatrix}
(1+\mathbf{p}_{A}^2)(1-u)+u (1+\mathbf{p}_{A})&0&0&(1-\mathbf{p}_{A})(1-u)+u\sqrt{1-\mathbf{p}_{A}}\\
0&\mathbf{p}_{A}(1-\mathbf{p}_{A})(1-u)&0&0\\
0&0&\mathbf{p}_{A}(1-\mathbf{p}_{A})(1-u)&0\\
(1-\mathbf{p}_{A})(1-u)+u \sqrt{1-\mathbf{p}_{A}}&0&0&(1-u)(1-\mathbf{p}_{A})^2 + u(1-\mathbf{p}_{A})
\end{pmatrix}.
\end{eqnarray}
The expression of $\mathcal{N}^{(PT)}$ of the state $\hat{\mathcal{\varrho}}^{A}_{AB}$ is written as:
\begin{equation}\label{ADNEGA}
\mathcal{N}^{(PT)}_{Ad}= (1-\mathbf{p}_A)^2 (1-u) + u \sqrt{1-\mathbf{p}_A}.
\end{equation}
After performing the $BQT$ protocol using the quantum channel $\hat{\mathcal{\varrho}}_{AB}$, Alice and Bob get the  states:
\begin{equation}
\hat{\varrho}_{A}^{Ad}=\frac{1}{2}\Bigl(\mathbf{I}+\sum_{n=1,2,3}\alpha^{(Ad)}_n\hat{\gamma_n}\Bigr), ~ \hat{\varrho}_{B}^{Ad}=\frac{1}{2}\Bigl(\mathbf{I}+\sum_{n=1,2,3}\beta^{(Ad)}_n \hat{\delta_n}\Bigr),
\end{equation}
where,
\begin{eqnarray}\label{100}
\alpha_{x}^{(Ad)}&=&{ \mathcal{M_A}} \bar{ \mathcal{M_B}} (\mathcal{N}^{(PT)}_{Ad}-(1-u)(\mathbf{p}_{A}^2-\mathbf{p}_{A})) \cos\varphi_A  \sin\Bigr(\vartheta_A\Bigl),\nonumber\\
\alpha_{y}^{(Ad)}&=& - { \mathcal{M_A}} \bar{ \mathcal{M_B}} (\mathcal{N}^{(PT)}_{Ad}-(1-u)(\mathbf{p}_{A}^2-\mathbf{p}_{A}))  \sin\varphi_A \sin\Bigr(\vartheta_A\Bigl),\nonumber\\
\alpha_{z}^{(Ad)}&=&{ \mathcal{M_A}} \bar{ \mathcal{M_B}} \Bigr( \mathbf{A} \cos\Bigr(\frac{\vartheta_A}{2}\Bigl)^2 + \mathbf{B} \sin\Bigr(\frac{\vartheta_A}{2}\Bigl)^2\Bigl) + \mathbf{p}_{A}(1-{ \mathcal{M_A}} \bar{ \mathcal{M_B}}),
\end{eqnarray} 
and
\begin{eqnarray}
\beta_{x}^{(Ad)}&=&{ \mathcal{M_B}} \bar{ \mathcal{M_A}}(\mathcal{N}^{(PT)}_{Ad}-(1-u)(\mathbf{p}_{A}^2-\mathbf{p}_{A})) \cos\varphi_B  \sin\Bigr(\vartheta_B\Bigl),\nonumber\\
\beta_{y}^{(Ad)}&=&- { \mathcal{M_B}} \bar{ \mathcal{M_A}} (\mathcal{N}^{(PT)}_{Ad}-(1-u)(\mathbf{p}_{A}^2-\mathbf{p}_{A})) \sin\varphi_B \sin\Bigr(\vartheta_B\Bigl),\nonumber\\
\beta_{z}^{(Ad)}&=&{ \mathcal{M_B}} \bar{ \mathcal{M_A}} \Bigr( \mathbf{A} \cos\Bigr(\frac{\vartheta_B}{2}\Bigl)^2 + \mathbf{B} \sin\Bigr(\frac{\vartheta_B}{2}\Bigl)^2\Bigl) + \mathbf{p}_{A}(1-{ \mathcal{M_B}} \bar{ \mathcal{M_A}}),
\end{eqnarray}
whereas:
\begin{eqnarray}
\mathbf{A}&=&(1-\mathbf{p}_{A})(1-2\mathbf{p}_{A}+2u\mathbf{p}_{A}); \quad\quad \mathbf{B}=(1-u)(\mathbf{p}_{A}-2\mathbf{p}_{A}^2 -1) - u(1+\mathbf{p}_{A}).
\end{eqnarray}
The teleportation fidelities are given as: 
\begin{eqnarray}
    f_{Ad}^{A \rightarrow B} = { \mathcal{M_A}} \bar{ \mathcal{M_B}} \bigl(\chi \sin(\vartheta_A)^2 +\delta \cos(\frac{\vartheta_A}{2})^4+\eta \sin(\frac{\vartheta_A}{2})^4 \bigr) + \frac{1-{ \mathcal{M_A}} \bar{ \mathcal{M_B}}}{2}(1+\mathbf{p}_A \cos(\vartheta_A)); \nonumber \\
       f_{Ad}^{B \rightarrow A} =  { \mathcal{M_B}} \bar{ \mathcal{M_A}} \bigl(\chi \sin(\vartheta_B)^2 +\delta \cos(\frac{\vartheta_B}{2})^4+\eta \sin(\frac{\vartheta_B}{2})^4 \bigr) + \frac{1-{ \mathcal{M_B}} \bar{ \mathcal{M_A}}}{2}(1+\mathbf{p}_A \cos(\vartheta_B)).
\end{eqnarray}
With different elements presented as
\begin{eqnarray}
\chi=\mathcal{N}^{(PT)}_{Ad}-(1-u)(2\mathbf{p}_A^2-2\mathbf{p}_A);\quad\quad \delta=\mathcal{N}^{(PT)}_{Ad}+u(1-\mathbf{p}_A-\sqrt{1-\mathbf{p}_A});\quad\quad \eta=\mathcal{N}^{(PT)}_{Ad}+(u+2\mathbf{p}_A-u\sqrt{1-\mathbf{p}_A}).
\end{eqnarray}
To explore the $\mathbf{p}_A$ time dependence, we consider a well studied model that describes the interaction of the two qubits. Each one interacts only and independently with its local environment formed by the quantized modes of a high-Q cavity. The Hamiltonian is given as \cite{AD0}:
\begin{equation}
\hat{H}=\omega_0 \sigma_{+} \sigma_{-} + \sum_k \omega_k r^{\dagger}_k r_k + (\sigma_{+} B + \sigma_{-} B^{\dagger}),
\end{equation}
where $B=\sum_{k}g_k r_k $, $\omega_0$ is the transition frequency of the two-level qubit, $\sigma_{\pm}$ are the lowering and raising operators and  $k$ stands for the field modes of the reservoir with frequencies $\omega_k$. Besides, $r_k , r^{\dagger}_k$ are the modes creation and annihilation operators while $g_k$ are the coupling constants.  Indeed, two coupling regimes can be distinguished; namely weak and strong regimes. In case of a weak regime, the relaxation time is greater than the reservoir correlations time $\gamma < \Gamma /2$, $\tau_R > 2\tau_B$. The decoherence factor is, therefore, given as follows \cite{12}:
\begin{equation}\label{deco}
\mathbf{p}(t)_{A}=e^{-\Gamma t}\Bigr(\cosh(\frac{dt}{2})+ \frac{\Gamma}{d} \sinh(\frac{dt}{2})\Bigl)^2,
\end{equation}
whereas, $d = i \sqrt{2\gamma \Gamma - \Gamma^2}$.  $\Gamma$ depends on the reservoir correlation time $\tau_r \approx \Gamma^{-1}$ and $\gamma$ is the coupling strength related to qubit relaxation time $\tau_q \approx \gamma^{-1}$.
\\

Besides, in the strong coupling regime, for $\gamma > \Gamma /2$, $\tau_R < 2\tau_B$,  the large reservoir correlation time  is greater or of the same order of the relaxation time. Consequently, the non-Markovian effects become important. In the strong coupling regime, the decoherence factor can be obtained by substituting the hyperbolic functions with the corresponding harmonic ones and $d=\sqrt{2\gamma \Gamma - \Gamma^2}$ in Eq.(\ref{deco}).

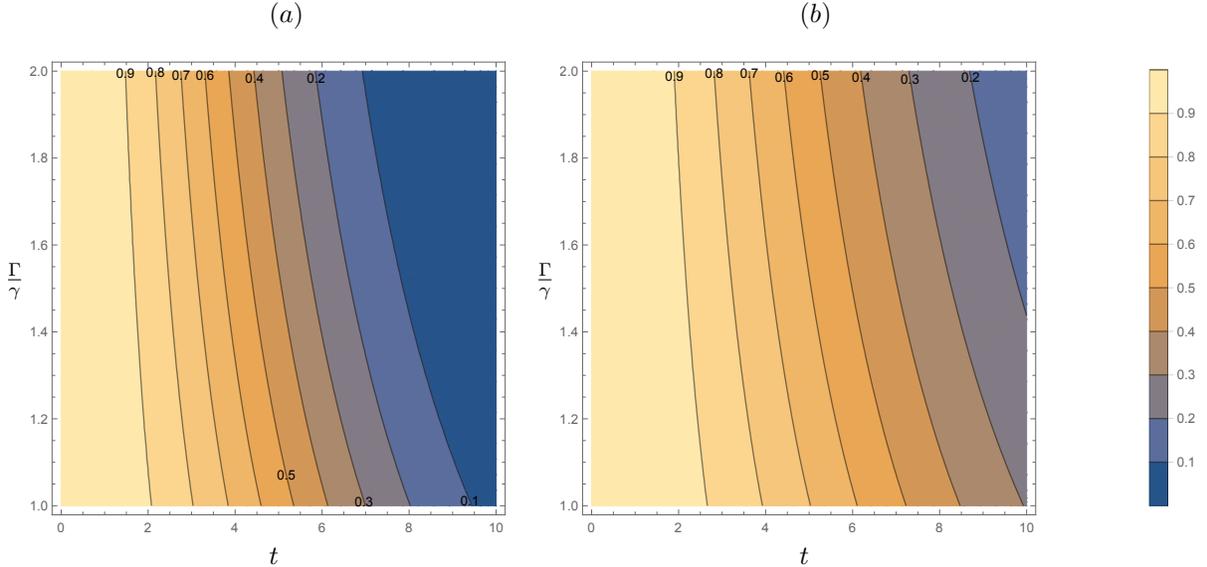
\begin{figure}[!htb]
\begin{center}
\includegraphics[scale=.5]{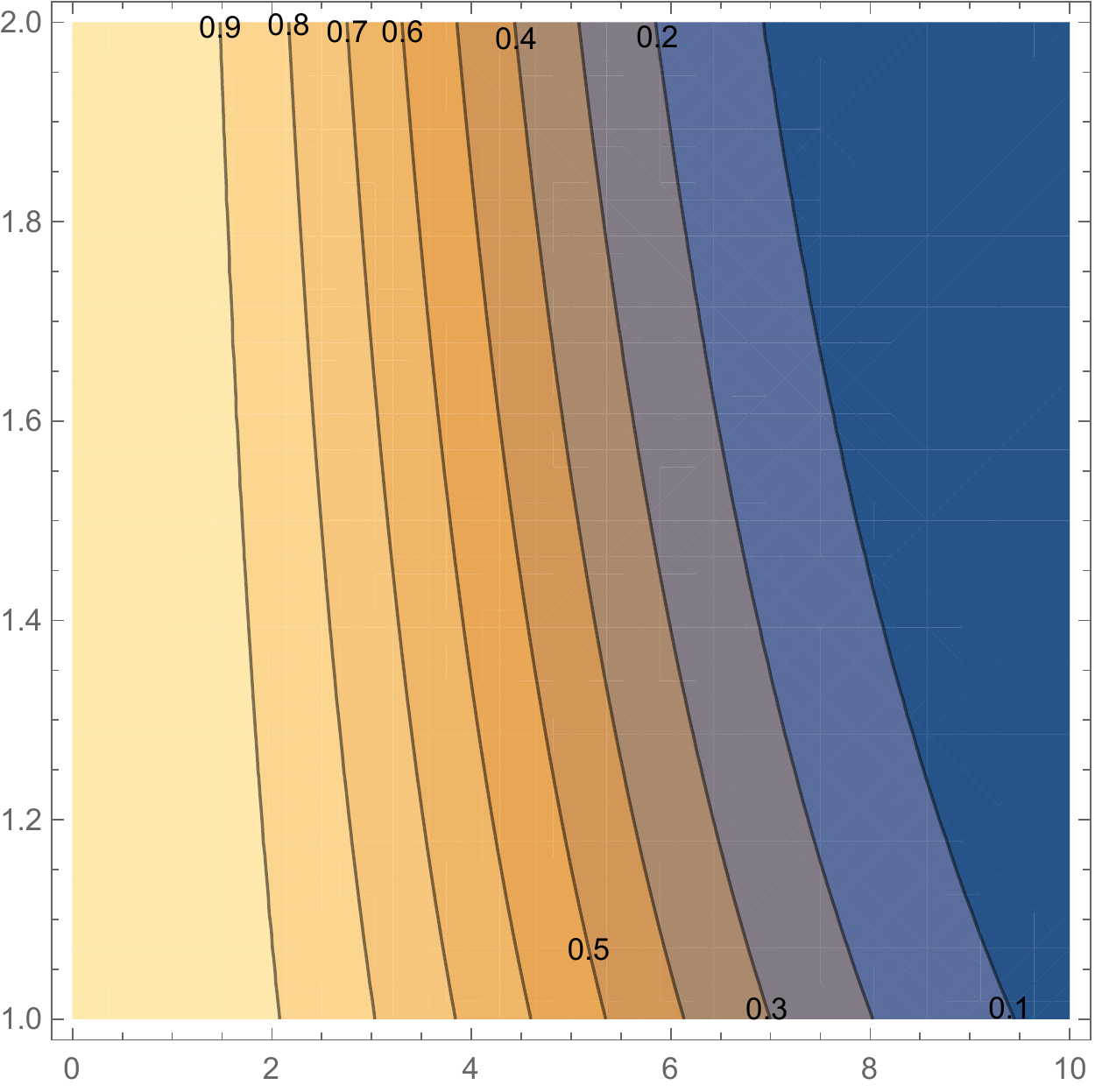}
\put(-190,95){$\frac{\Gamma}{\gamma}$}
                      \put(-90,195){$(a)$}
                         \put(-90,-10){$t$}\quad\quad
\includegraphics[scale=.5]{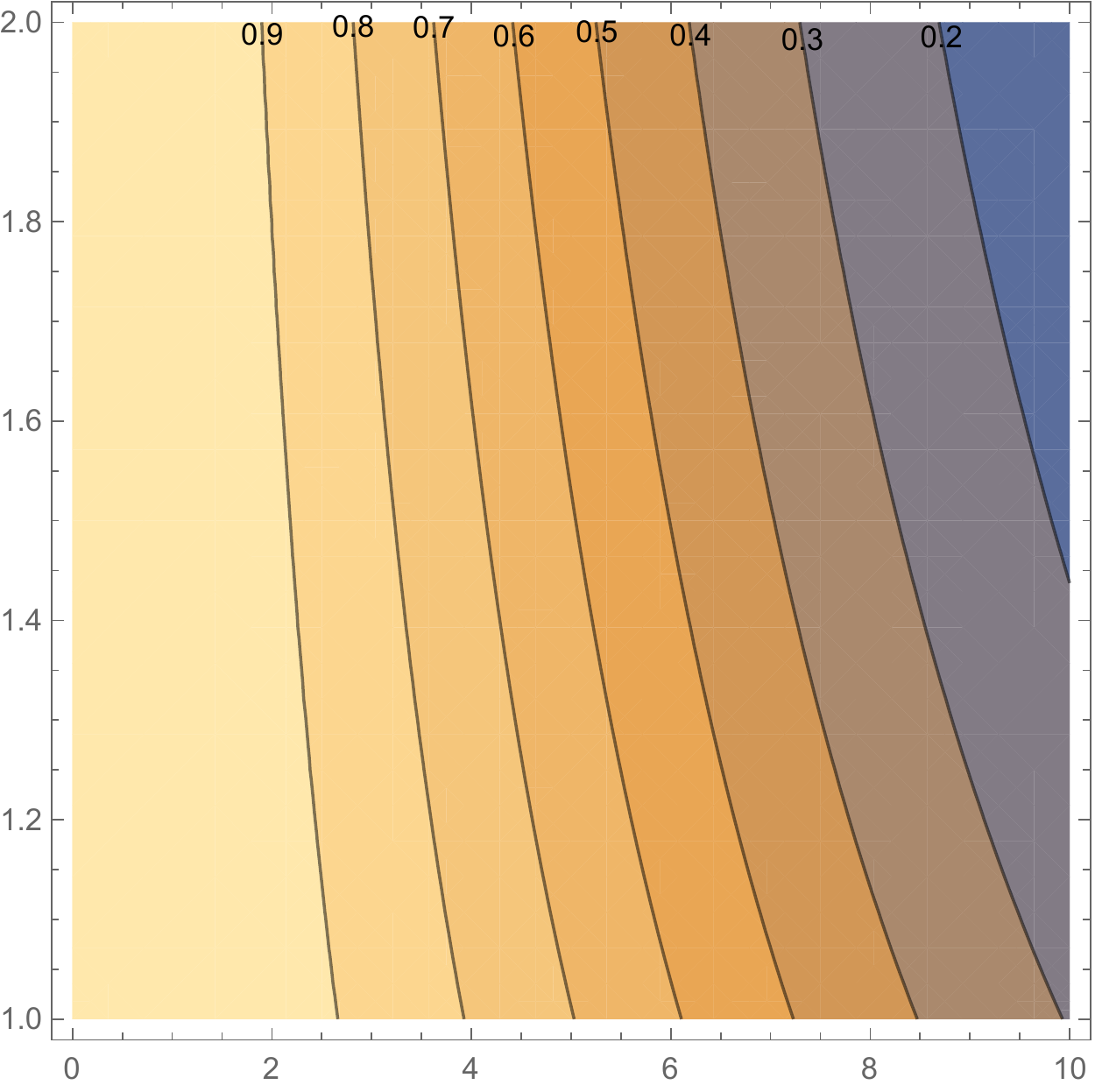}
\put(-190,95){$\frac{\Gamma}{\gamma}$}
                      \put(-90,195){$(b)$}
                         \put(-90,-10){$t$}\quad\quad\quad\quad
\includegraphics[scale=.5]{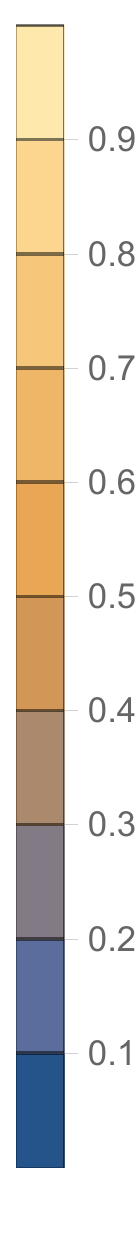}
\caption{Negativity of the partial transpose under Amplitude damping channel versus time in the Markovian regime ($\frac{\Gamma}{\gamma}>\frac{1}{2}$). (a): $u=0$, (b): $u=0.5$}
\label{FIGURE333}
 \end{center}
\end{figure}
\begin{figure}[!htb]
\begin{center}
\includegraphics[scale=.6]{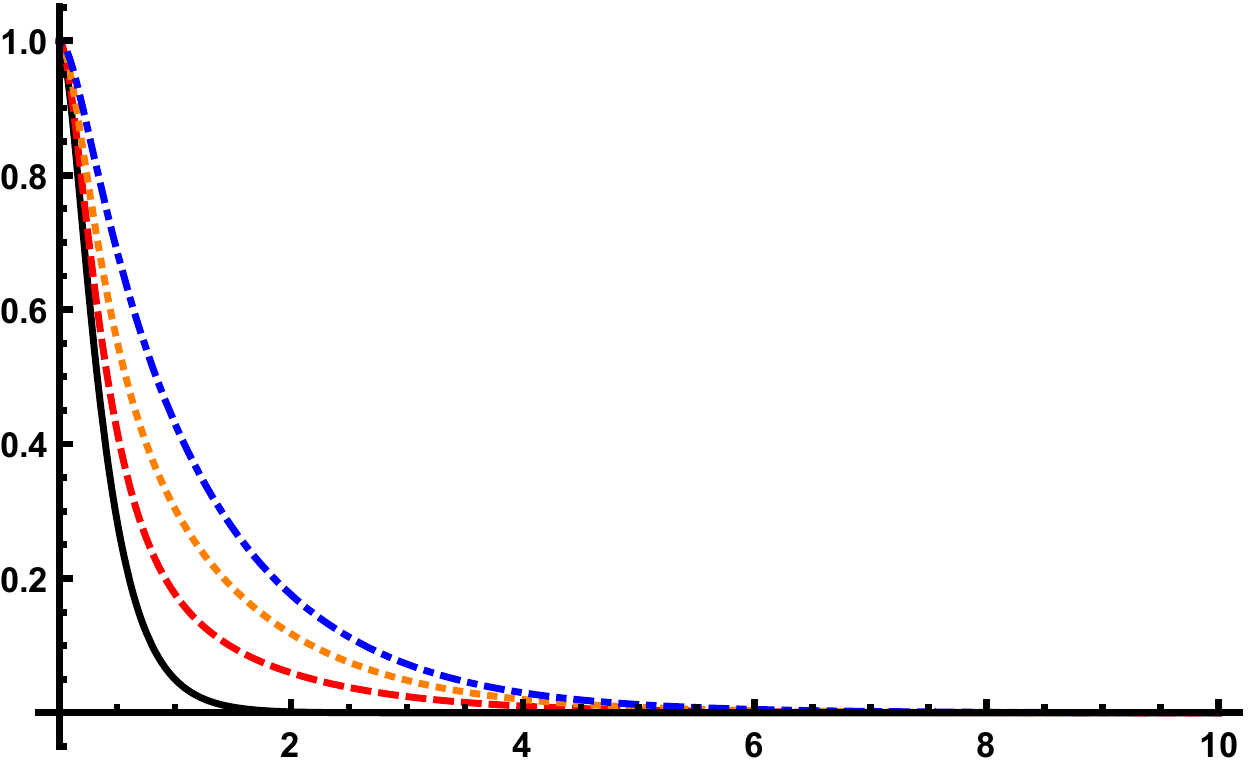}
\put(-255,95){$\mathcal{N}^{(PT)}_{Ad}$}
                      \put(-100,110){$(a)$}
                         \put(-130,-5){$t^{*}$}\\
\includegraphics[scale=.6]{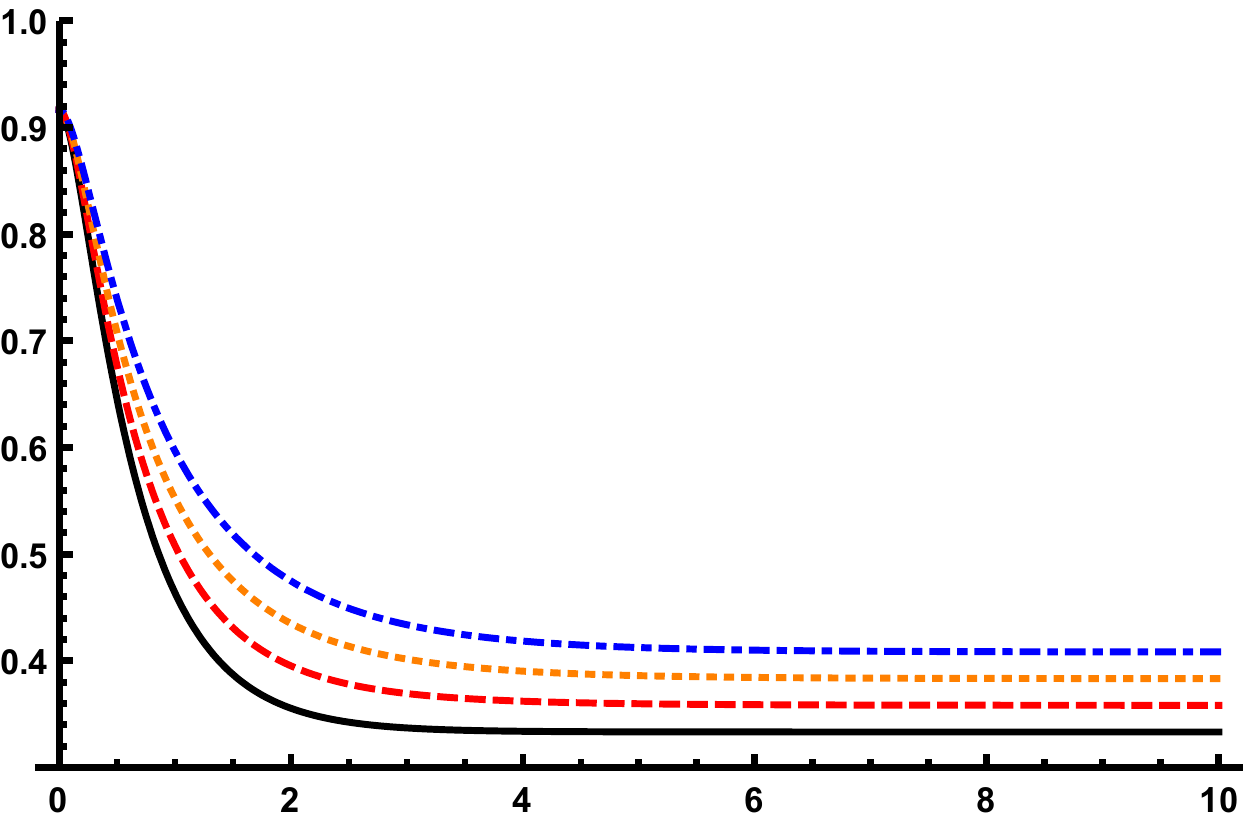}
\put(-255,95){$\mathcal{F}^{in\rightarrow out}_{Avg}$}
                      \put(-100,110){$(b)$}
                         \put(-130,-5){$t^{*}$}\quad\quad\quad
\includegraphics[scale=.6]{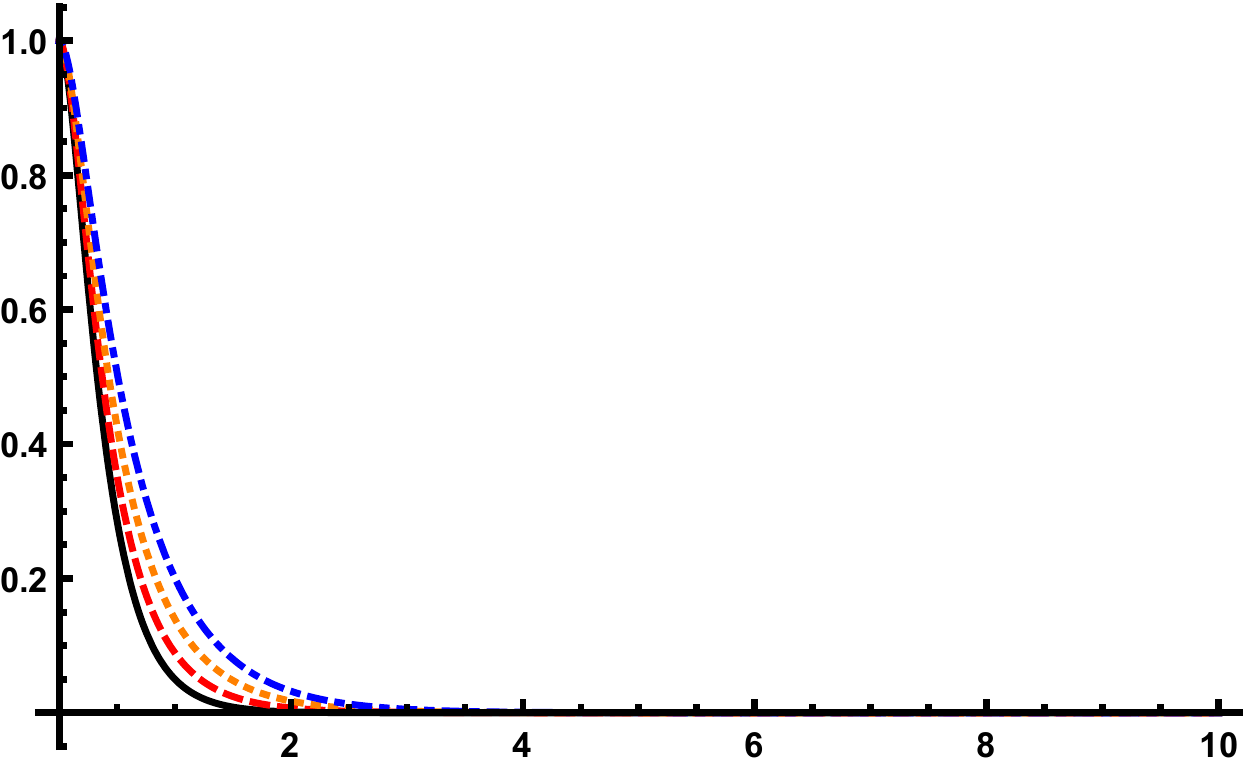}
\put(-255,95){$\mathcal{J}_{\vartheta_i}$}
                      \put(-100,110){$(c)$}
                         \put(-130,-5){$t^{*}$}
\caption{Negativity, average fidelity, and quantum Fisher information under Amplitude damping channel versus the re-scaled time in the Markovian regime ($\Gamma =5\gamma$). The solid black, red dashed, orange dotted and bue dotdashed correspond to $u=0, 0.3, 0.6$, and $0.9$, respectively. We set here $\vartheta_A=\vartheta_B=0$ and $\tilde\vartheta_A=0$, $\tilde\vartheta_B=\pi$.
}
\label{FIGURE3}
 \end{center}
\end{figure}

In Fig.\ref{FIGURE333}, we show the behavior of the negativity of the partial transpose $\mathcal{N}^{(PT)}_{Ad}$ of the quantum channel that is distributed through the amplitude damping channel in the Markovian regime, namely $\frac{\Gamma}{\gamma}>\frac{1}{2}$.  As a matter of fact, the amount of entanglement in the quantum channel is maximum for small value of $t$ and the bounds of the nagativity, $\mathcal{N}^{(PT)}_{Ad}$, decreases gradually with the flow of time.  However, as shown in Fig.\ref{FIGURE333}.(b), the exploitation of the classical correlations, where $u=0.5$, could slightly improve the amount of the entanglement for a larger period of the time evolution. In order to more illustrate the behavior of the negativity, the average fidelities and the quantum Fisher information in the Markovian regime, we depict the three quantities behaviors, namely $\mathcal{N}^{(PT)}_{Ad}$, $\mathcal{F}^{in\rightarrow out}_{Avg}$, $\mathcal{J}_{\vartheta_i}$ in Figs \ref{FIGURE3}(a,b,c). As shown in these figures, In the Markovian regime of the amplitude damping channel we have  $\gamma < \Gamma/2$ and $d=-\Gamma$, the three quantities are decreasing monotonically with time and frozen with values $0, 0.34$ and $0$, respectively. However, by harnessing the classical correlations between the noise actions one can slightly increase the three quantities, where the three quantities, $\mathcal{N}^{(PT)}_{Ad}$, $\mathcal{F}^{in\rightarrow out}_{Avg}$, $\mathcal{J}_{\vartheta_i}$ are frozen with values $0, 0.3+\frac{u}{2n},0$, respectively, where $n=1,2,3,4$.

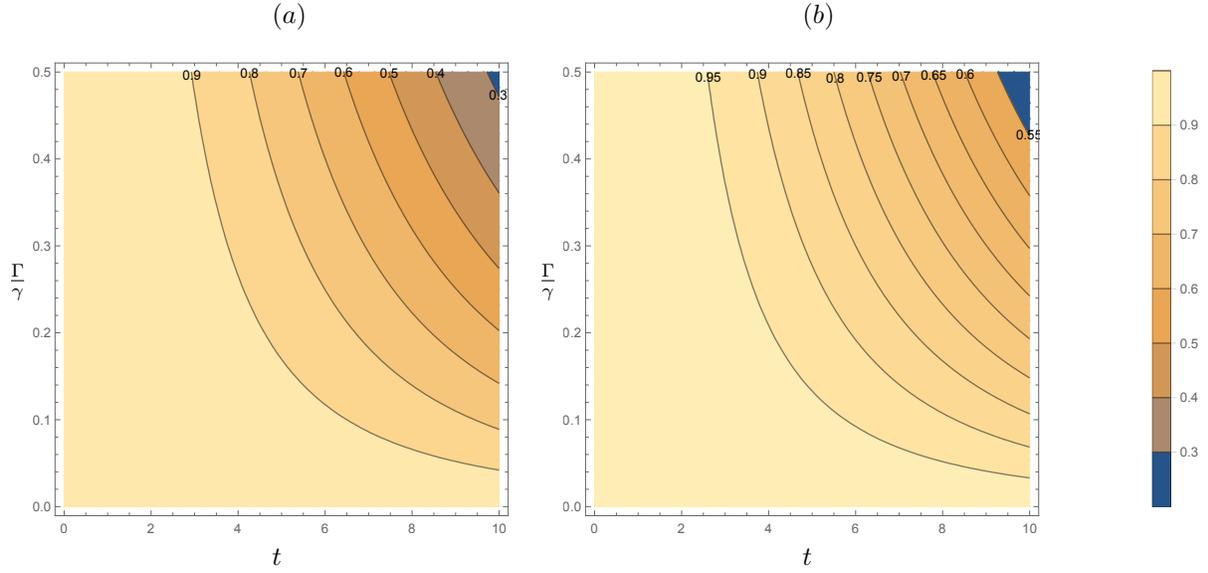
\begin{figure}[!htb]
\begin{center}
\includegraphics[scale=.5]{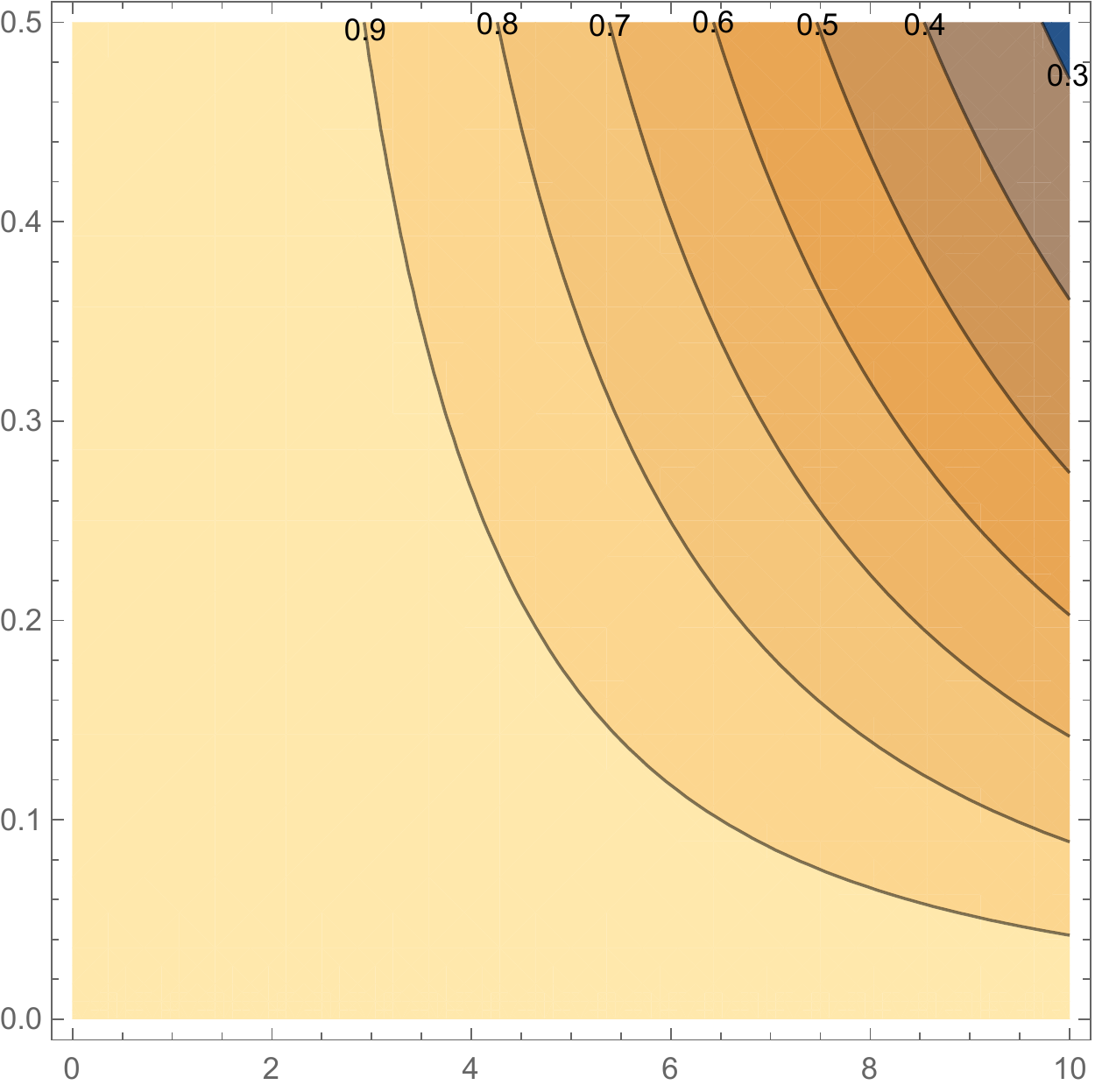}
\put(-190,95){$\frac{\Gamma}{\gamma}$}
                      \put(-90,195){$(a)$}
                         \put(-90,-10){$t$}\quad\quad
\includegraphics[scale=.5]{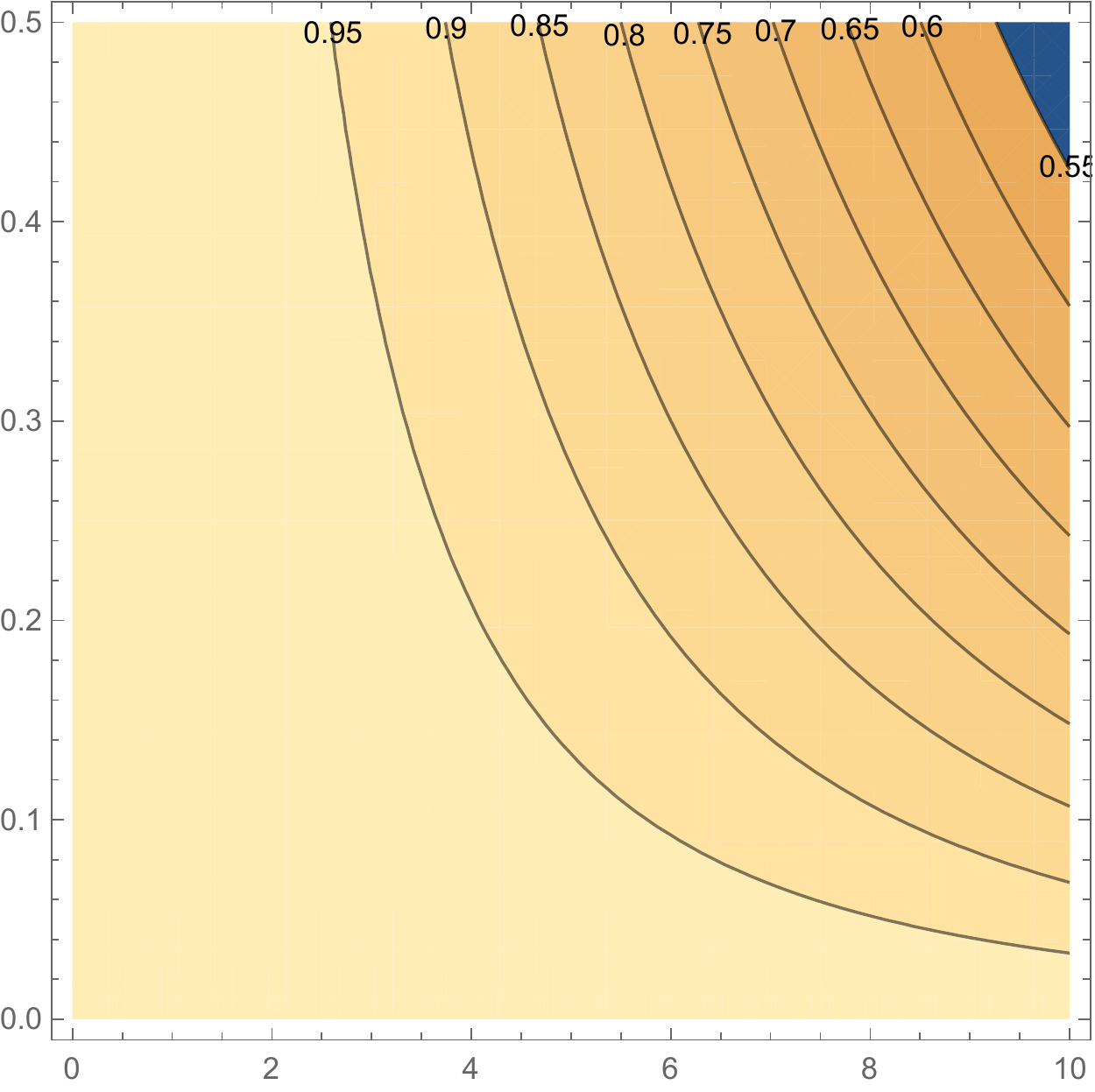}
\put(-190,95){$\frac{\Gamma}{\gamma}$}
                      \put(-90,195){$(b)$}
                         \put(-90,-10){$t$}\quad\quad\quad\quad
\includegraphics[scale=.5]{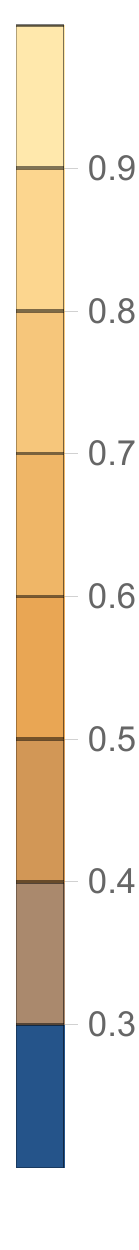}
\caption{Negativity of the partial transpose under Amplitude damping channel versus time in the non- Markovian regime ($\frac{\Gamma}{\gamma}<\frac{1}{2}$). (a): $u=0$, (b): $u=0.5$
}
\label{FIGUREx}
 \end{center}
\end{figure}
In the non-Markovian regime, we show the behavior of the entanglement under the amplitude damping channel in Fig.\ref{FIGUREx}.  The negativity $\mathcal{N}^{(PT)}_{Ad}$ could resist more with the flow of time compared to the behavior shown in Fig.\ref{FIGURE333}. Similarly, by taking into account the amplitude damping channel's correlations, one can improve the entanglement at a later period of time.\\

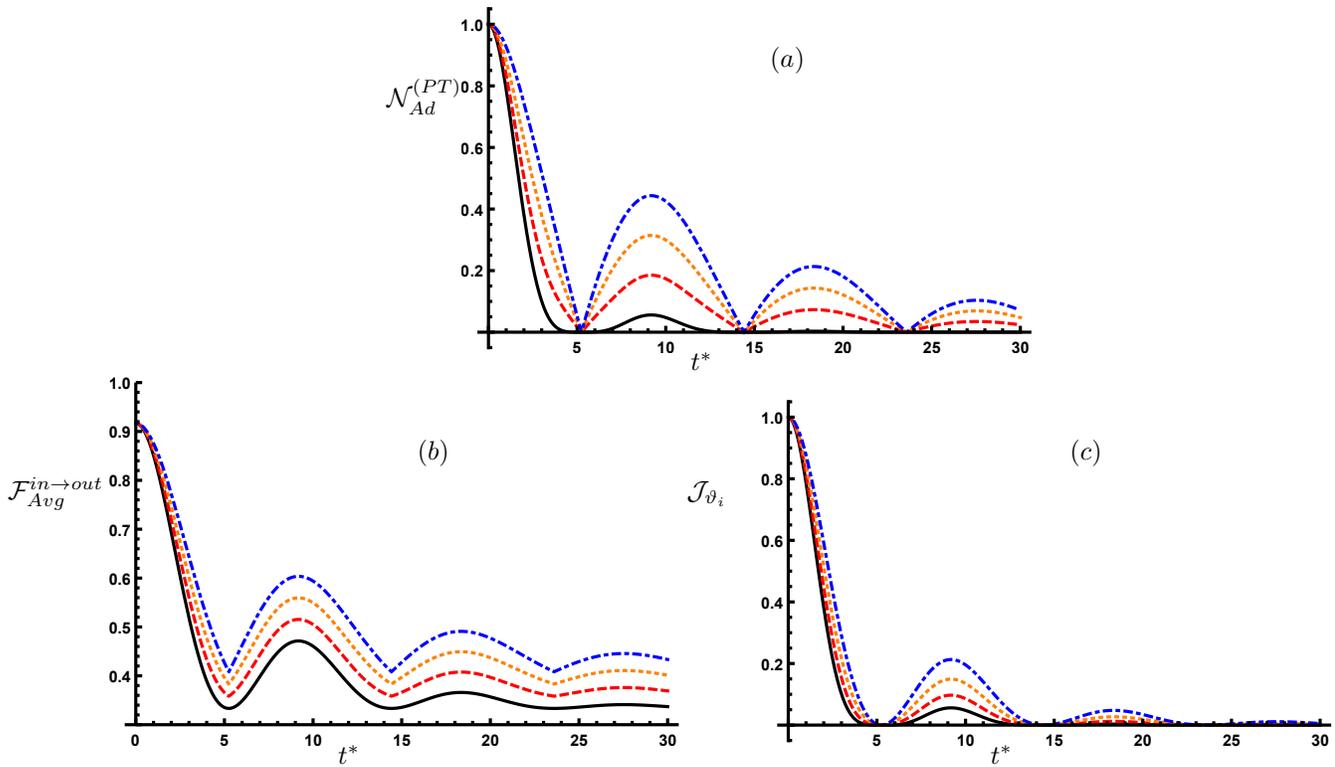
\begin{figure}[!htb]
\begin{center}
\includegraphics[scale=.6]{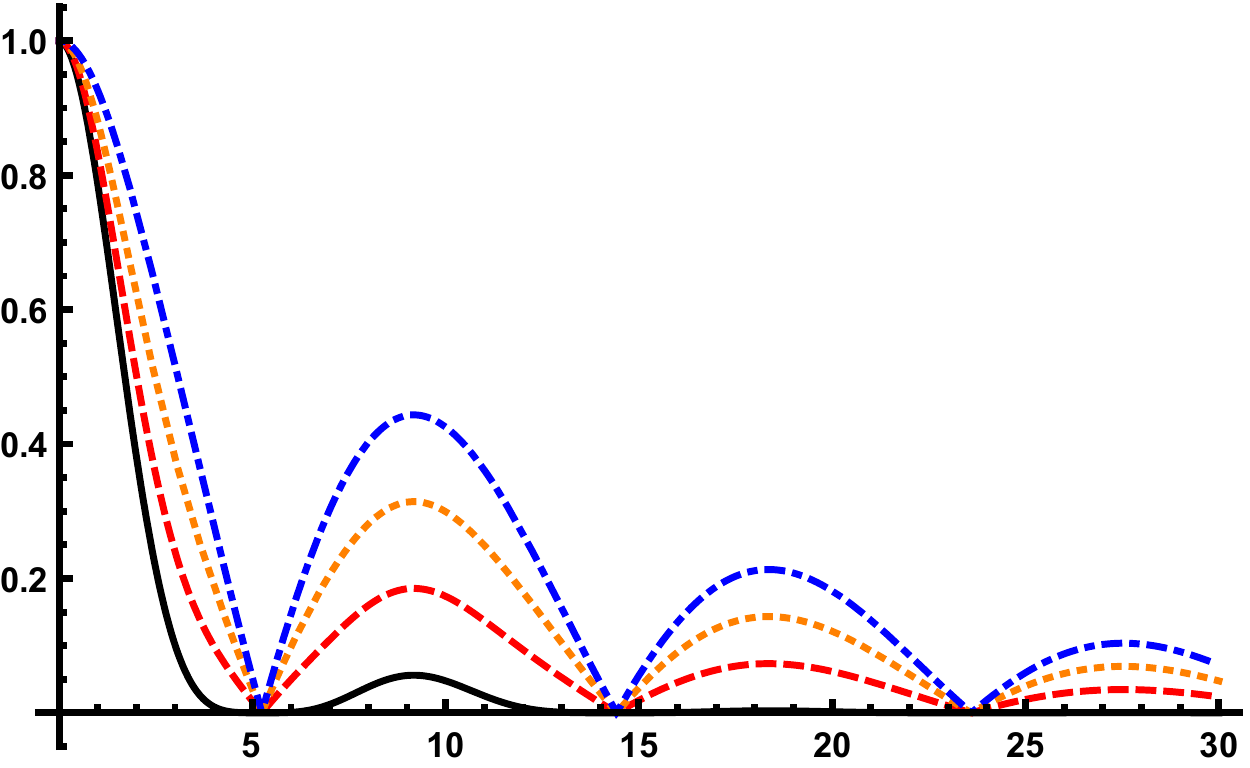}
\put(-245,95){$\mathcal{N}^{(PT)}_{Ad}$}
                      \put(-100,110){$(a)$}
                         \put(-130,-5){$t^{*}$}\\
\includegraphics[scale=.6]{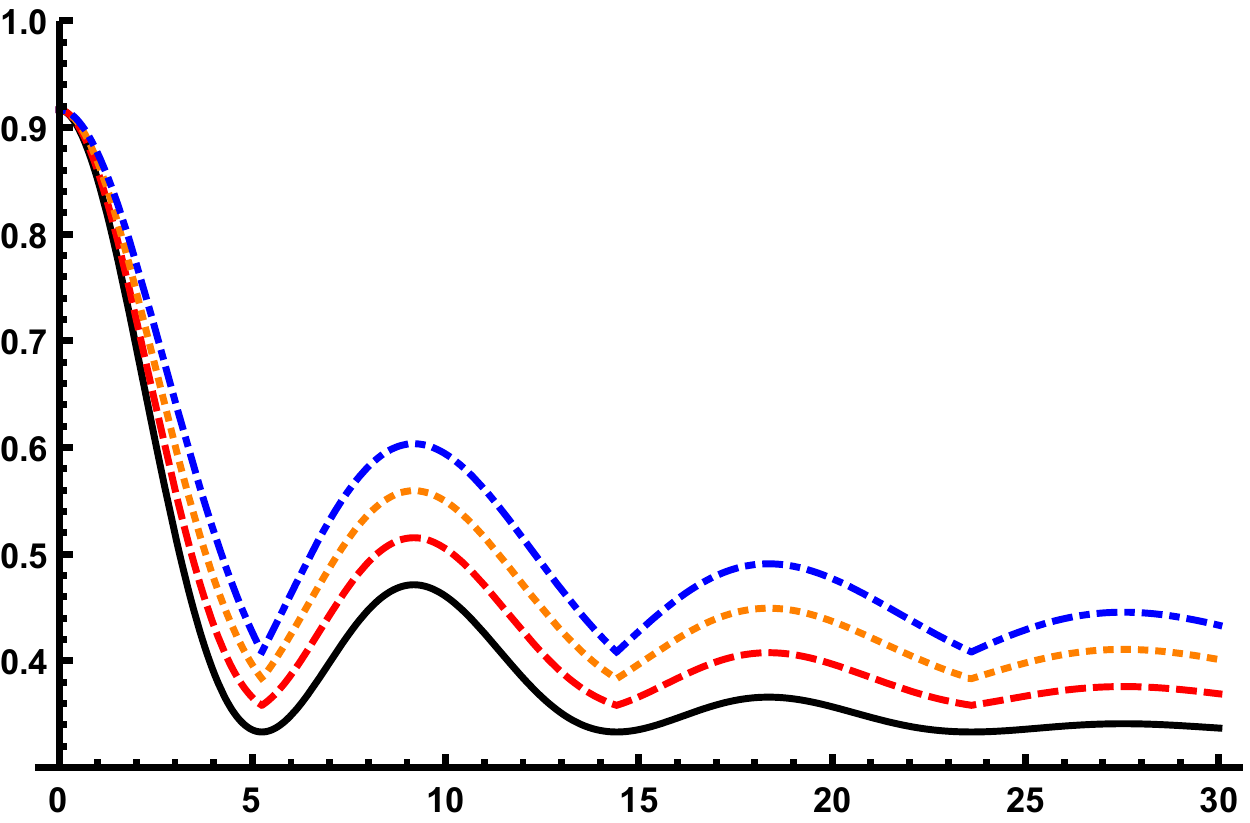}
\put(-255,95){$\mathcal{F}^{in\rightarrow out}_{Avg}$}
                      \put(-100,110){$(b)$}
                         \put(-130,-5){$t^{*}$}\quad\quad\quad
\includegraphics[scale=.6]{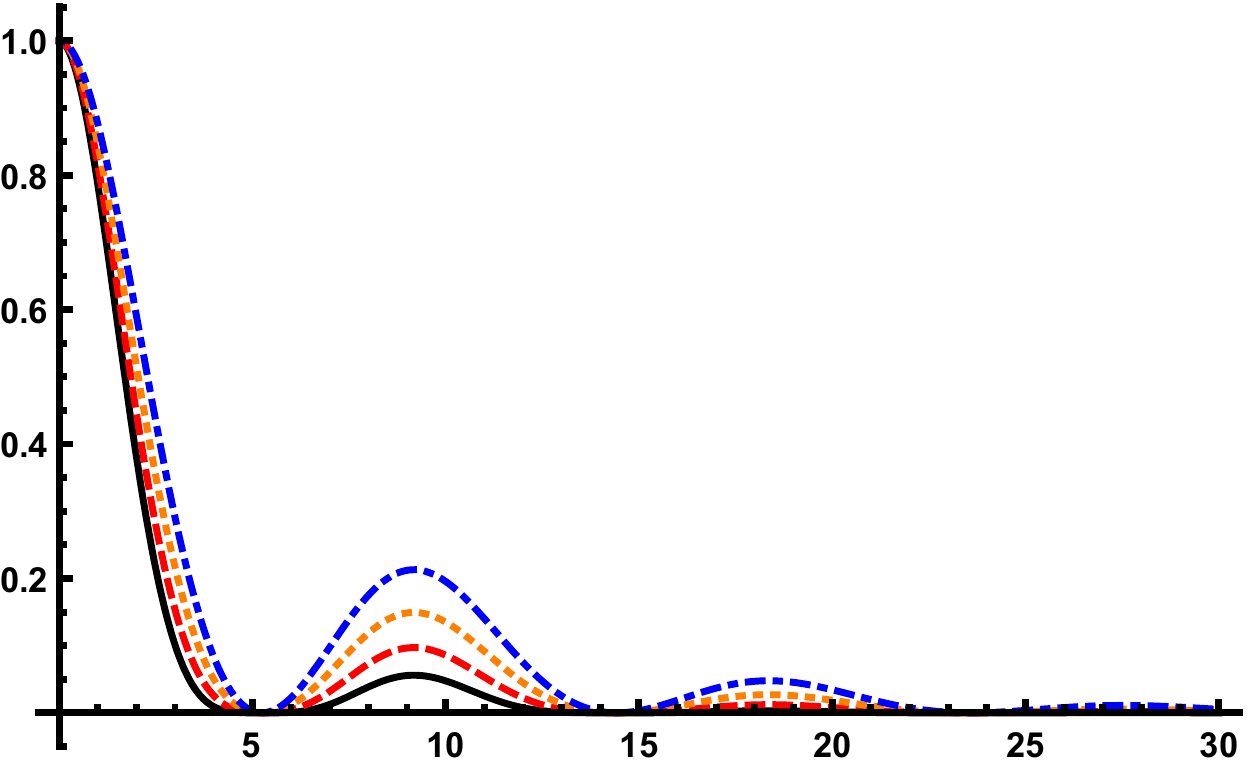}
\put(-245,95){$\mathcal{J}_{\vartheta_i}$}
                      \put(-100,110){$(c)$}
                         \put(-130,-5){$t^{*}$}\quad\quad\quad
\caption{Negativity, average fidelity, and quantum Fisher information versus the re-scaled time in non-Markovian regime with $\Gamma =0.1\gamma$. The solid black, red dashed, orange dotted and blue dotdashed correspond to $u=0, 0.3, 0.6$, and $0.9$, respectively. We set here $\vartheta_A=\vartheta_B=0$ and $\tilde\vartheta_A=0$, $\tilde\vartheta_B=\pi$.}
\label{FIGURE4}
 \end{center}
\end{figure}

In Fig.\ref{FIGURE4}, we exhibit the behavior of the negativity, average fidelity, the quantum Fisher information for the parameter $\vartheta$, namely $\mathcal{N}^{(PT)}_{Ad}$, $\mathcal{F}^{in\rightarrow out}_{Avg}$, $\mathcal{J}_{\vartheta_i}$ in the non-Markovian regime of the correlated amplitude damping channel; where $\gamma > \Gamma/2$ and $d=\sqrt{2\gamma \Gamma}$, we have $\mathbf{p}(t)_{A}=e^{-\Gamma t}\cos(dt/2)^2$. Indeed, the damped oscillations have a period of time which is $T=9\pi$ of their evolution against time t. It is clear that the three quantities have maximum peaks at the instants of time $t=nT$ with $n\in\mathbb{N}$; these peaks decrease with time flow. Furthermore, at the instant time $t=n\frac{T}{2}$with $n\in \mathbb{N}^{*}$ the tree quantities decrease to their minimum values $0, 0.3+\frac{u}{2n}, 0$, respectively.

\newpage
\section{Summary}

In this work, we have investigated the combined memory effects, namely the classical and temporal correlations arising in noisy quantum channels,  on the bidirectional quantum teleportation of single-qubit states between two parts, namely $Alice$ and $Bob$. Firstly, we examined the quantum channel entanglement of the bidirectional quantum teleportation protocol, using the negativity of the partial transposed density matrix, $\mathcal{N}^{(PT)}$, as an entanglement quantifier. We have derived $\mathcal{N}^{(PT)}$ of the quantum channel that is distributed through typical decoherence channels such as the dephasing and amplitude damping channels. The expressions of the negativity, $\mathcal{N}^{(PT)}$, depend on the decoherence strength and on the degree of the correlations in the considered noise channel. We displayed that $\mathcal{N}^{(PT)}$ is slightly increased if one takes into account the classical correlations between the noisy channel actions in the Markovian regime. However, $\mathcal{N}^{(PT)}$ could be further enhanced  by considering the correlations of the channel in the non-Markovian regime.\\

Furthermore, we have explored the teleportation average fidelities and the quantum Fisher information of the weight parameters, $\vartheta_A$ and $\vartheta_B$. The explicit forms of the teleportation fidelities $f^{in \rightarrow out}$, and the quantum Fisher information $\mathcal{J}_{\vartheta_i}$ (with $i=A,B$) are obtained for the dephasing and the amplitude damping channels. The expression of the teleportation fidelities and the quantum Fisher information depend on the amount of the survival amount of entanglement in the $BQT$ quantum channel,  the decoherence factor, the degree of the correlation of the noise channels, and also on all the initial state settings, namely ($\vartheta_i, \varphi_i, \tilde\vartheta_i$, for $ i=A,B$) . It is demonstrated that  both quantities $\mathcal{F}^{in \rightarrow out}_{Avg}$ and $\mathcal{J}_{\vartheta_i}$ are decreasing monotonically in the course of the time evolution in the Markovian regime. However, by considering the classical correlations between the successive actions of the decoherence channels on the $BQT$ quantum channel, one could slightly increase these quantities. Besides, we have discussed the effects of the temporal channel correlations in the non-Markovian regime on the average teleportation fidelities, and on the quantum Fisher information. We have shown that both quantities could be improved due to the temporal correlations of the noisy channels. Moreover, the drastic effect of noise that decreases the average teleportation fidelities and the quantum Fisher information, with time flow, could be mitigated for a large range of time by taking into account the combined effect of the classical and temporal correlations between the successive actions of the noisy channels.\\

$In$ $conclusion$: The effect of  the dephasing and the amplitude damping noise channels causes a deterioration of entanglement, and a monotonic decrease in both the average teleportation fidelities  and the quantum Fisher information. This deterioration of entanglement and the decay of the average teleportation fidelities and quantum Fisher information could be relatively recovered as long as possible, by considering the combined effects of the noise channel correlations.


\begin{thebibliography}{9}
\bibitem{1}C. H. Bennett, G.  Brassard, C. Crépeau, R. Jozsa, A. Peres, W. K.  Wootters, Teleporting an unknown quantum state via dual classical and Einstein-Podolsky-Rosen channels, Physical reviewletters 70(13) p 1895 (1993).
\bibitem{a}S. Pirandola, J. Eisert, C. Weedbrook, A. Furusawa, S. L. Braunstein, Advances in quantum teleportation. Nature photonics, 9(10), 641-652 (2015).
\bibitem{2}H. J. Kimble, Nature (London) 453, 1023 (2008).
\bibitem{allati1}A. El Allati, Y. Hassouni and N. Metwally, Communication via entangled coherent quantum Network, Phys. Scr. 83, 065002 (2011).
\bibitem{3}H.J. Briegel, W. Dur, J.I. Cirac, and P. Zoller, Phys. Rev. Lett. 81, 5932 (1998).
\bibitem{4}D. Gottesman and I.L. Chuang, Nature (London) 402, 390 (1999).
\bibitem{b}D. Bouwmeester, J. W Pan,  K. Mattle, M. Eibl, H. Weinfurter, A. Zeilinger, Experimental quantum teleportation. Nature, 390(6660), 575-579 (1997).
\bibitem{c}D. Llewellyn, Y. Ding,  I. I. Faruque, S. Paesani, D. Bacco, R. Santagati, M. G. Thompson, Chip-to-chip quantum teleportation and multi-photon entanglement in silicon. Nature Physics, 16(2), 148-153  (2020).
\bibitem{Lev0} L. Vaidman,  N Erez, A. Retzker,  Another Look at Quantum Teleportation. International Journal of Quantum Information, 4(01), 197-208 (2006).
\bibitem{Lev}L. Vaidman, Teleportation of quantum states. Physical Review A, 49(2), 1473 (1994).
\bibitem{5}E. O. Kiktenko, A. A. Popov,  A. K. Fedorov, Bidirectional imperfect quantum teleportation with a single Bell state. Physical Review A, 93(6), 062305 (2016).
\bibitem{6}X.-W. Zha, Z.-C. Zou, J.-X. Qi and H.-Y. Song, Int. J. Theor. Phys, 52, 1740 (2013); C. Shukla, A. Banerjee and A. Pathak, Int. J. Theor. Phys, 52, 3790 (2013); Y.-H. Li, X.-I. Li, M.-H. Sang, Y.-Y. Nie and Z.-S. Wang, Quantum Inf. Process, 12, 3835 (2013); Y. Chen, Int. J. Theor. Phys, 53, 1454 (2014).
\bibitem{7}C. Seida, A. El Allati,  N. Metwally, Y. Hassouni,  Multi-party bidirectional teleportation. Optik, 167784  (2021).
\bibitem{7a}R. Jozsa, Fidelity for mixed quantum states. Journal of modern optics, 41(12), 2315-2323  (1994).
\bibitem{7b}M. G. Paris, Quantum estimation for quantum technology. International Journal of Quantum Information, 7(supp01), 125-137   (2009).
\bibitem{d}C. Seida, A. El Allati, N. Metwally, Y. Hassouni,  Bidirectional teleportation using Fisher information. Modern Physics Letters A, 35(33), 2050272 (2020).
\bibitem{9}S. L. Braunstein and C. M. Caves, Phys. Rev. Lett. 72, 3439 (1994).
\bibitem{10}M. G. Paris, Quantum estimation for quantum technology. International Journal of Quantum Information, 7(supp01), 125-137  (2009).
\bibitem{111}D. DiVincenzo, B. Terhal,  Decoherence: the obstacle to quantum computation. Physics world, 11(3), 53. (1998).
\bibitem{0}S. Oh,  S. Lee, H. W. Lee,  Fidelity of quantum teleportation through noisy channels. Physical Review A, 66(2), 022316 (2002).
\bibitem{00}O. Gühne, F. Bodoky, M. Blaauboer,   Multiparticle entanglement under the influence of decoherence. Physical Review A, 78(6), 060301 (2008).
\bibitem{000}Z. Huang, H. Situ, Optimal protection of quantum coherence in noisy environment. International Journal of Theoretical Physics, 56(2), 503-513 (2017).
\bibitem{12}H. P. Breuer, F. Petruccione, The theory of open quantum systems. Oxford University Press on Demand (2002).
\bibitem{13}J. W. Pan, C. Simon, C. Brukner, Entanglement purification for quantum communication, Nature vol. 410 no 6832, p.
1067-1070 (2001).
\bibitem{14} P. G. Kwait, A. J. Berglund, J. B Altepeter, Experimental verification of decoherence-free subspaces, Science vol.
428 290 no 5491, p. 498-501 (2000).
\bibitem{15} S. Maniscalo, S. Francica, F. Zaffino, L. Rosa, Protecting entanglement via the quantum Zeno effect, Physical review
430 letters vol. 100, no 9, p. 090503 (2008).
\bibitem{16} 25. Y. S. Kim, J. C. Lee, O. Kwon, Protecting entanglement from decoherence using weak measurement and quantum
 measurement reversal, Nature Physics vol. 8 no 2, p. 117-120 (2012).
\bibitem{17} A. N. Korotkov , K. Kyle, Decoherence suppression by quantum measurement reversal, Physical Review A 81 4,
434 040103 (2010).
\bibitem{177} G. Lindblad, On the generators of quantum dynamical semigroups. Commun. Math. Phys. 48, 119 (1976).
\bibitem{18}H.-P. Breuer, E.-M. Laine, J. Piilo, Phys. Rev. Lett. 103, 210401 (2009).
\bibitem{19} E.-M. Laine, J. Piilo, and H.-P. Breuer, Phys. Rev. A 81, 062115 (2010).
\bibitem{20} F. Caruso, V. Giovannetti, C. Lupo, and S. Mancini, Rev. Mod. Phys. 86, 1203, 2014.;  H. P. Breuer and F. Petruccione, The Theory of Open Quantum Systems (Oxford: Oxford University Press) (2006). 
\bibitem{199}S. B. Xue, R. B. Wu, W. M. Zhang, J.  Zhang, C. W. Li,  T. J. Tarn, Decoherence suppression via non-Markovian coherent feedback control. Physical Review A, 86(5), 052304  (2012).
\bibitem{21}M. Hu, W. Zhou,  Enhancing two-qubit quantum coherence in a correlated dephasing channel. Laser Physics Letters, 16(4), 045201 (2019).
\bibitem{211}M. L. Hu, H. F. Wang, Protecting quantum Fisher information in correlated quantum channels. Annalen der Physik, 532(1), 1900378 (2020).
\bibitem{PD0}C. Seida,  A. El Allati, N. Metwally, Y. Hassouni, Bidirectional teleportation under correlated noise. The European Physical Journal D, 75(6), 1-12 (2021).
\bibitem{212} S. B. Xue, R. B. Wu , W. M. Zhang , J. Zhang , C. W. Li, T. J. Tarn   Phys. Rev. A 86 052304 (2012).
\bibitem{22}Y. N. Guo, K. Zeng, P. X. Chen, Teleportation of quantum Fisher information under decoherence channels with memory. Laser Physics Letters, 16(9), 095203  (2019).
\bibitem{23}Z. Y Wang, Z. Y. Qin,  Quantum teleportation, entanglement, and Bell nonlocality in correlated noisy channels. Laser Physics, 30(5), 055201 (2020).
\bibitem{24}C. Addis, G. Karpat,  C. Macchiavello, S. Maniscalco, Dynamical memory effects in correlated quantum channels. Physical Review A, 94(3), 032121 (2016).
\bibitem{241}Q. J. Tong, J. H. An, H. G. Luo, C. H. Oh, Decoherence suppression of a dissipative qubit by the non-Markovian effect. Journal of Physics B: Atomic, Molecular and Optical Physics, 43(15), 155501 (2010).
\bibitem{p1}C. Macchiavello and G. M. Palma, Phys. Rev. A 65, 050301 (2002).
\bibitem{Nega2}W. Son, M. S. Kim, et al. Dynamical entanglement transfer for quantum-information
networks. Physical Review A, vol. 70, no 2, p. 022320 (2004).

\bibitem{allati}A. El Allati, N. Metwally and Y. Hassouni, Transfer information remotely via noise entangled coherent channels.
Opt. Commun. 284, 519526 (2011).
\bibitem{444}W. Zhong, Z. Sun, J. Ma,  X. Wang, F. Nori,  Fisher information under decoherence in Bloch representation  (2013).
Physical Review A, 87(2), 022337.
\bibitem{PD1}M. M. Wilde, Quantum information theory. Cambridge University Press (2013).
\bibitem{p2}S. Daffer, K. Wodkiewicz, J. D. Cresser, J. K. McIver, Phys. Rev. A 70, 010304 (2004).
\bibitem{AD0}B. Bellomo, R. L. Franco,  G. Compagno,  Non-Markovian effects on the dynamics of entanglement. Physical Review Letters, 99(16), 160502 (2007).
\bibitem{AD1}D. A. Lidar. Lecture Notes on the Theory of Open Quantum Systems (2019).
\bibitem{AD2}A. Salles, F. de Melo, M. P. Almeida, M. Hor-Meyll, S. P. Walborn, P. H. Souto Ribeiro, and L. Davidovich. Experimental investigation of the dynamics of entanglement: Sudden death, complementarity, and continuous monitoring of the environment, Phys. Rev. A 78, 022322 (2008).
\bibitem{AD3}J. Preskill, Lecture Notes for Ph219/CS219: Quantum
Information (Caltech, 2018).
\bibitem{AD4}Y. Yeo, A. Skeen, Time-correlated quantum amplitude-damping channel. Physical Review A, 67(6), 064301 (2003).

\end{thebibliography}
\end{document}